\documentclass[sigplan,nonacm]{acmart}

\usepackage{enumitem}
\usepackage{xspace}
\usepackage{cleveref}
\usepackage{subcaption}
\usepackage{placeins}
\usepackage{makecell}
\usepackage{multirow}
\usepackage{tikz}
\usetikzlibrary{arrows.meta}
\usepackage{pgfplots}
\pgfplotsset{compat=1.3}
\usepackage{listings}
\usepgfplotslibrary{groupplots}

\setlist[description]{font=\normalfont\itshape\space}

\AtBeginDocument{%
  }

\usepackage{xcolor}
\definecolor{darkgreen}{rgb}{0.0, 0.5, 0.0}
\definecolor{codegreen}{rgb}{0,0.6,0}
\definecolor{codegray}{rgb}{0.5,0.5,0.5}
\definecolor{codepurple}{rgb}{0.58,0,0.82}
\definecolor{backcolour}{rgb}{0.95,0.95,0.92}
\definecolor{jcred}{HTML}{e31a1c}
\definecolor{jcgreen}{HTML}{33a02c}
\definecolor{jcblue}{HTML}{1f78b4}
\definecolor{jcorange}{HTML}{ff7f00}
\definecolor{jcpurple}{HTML}{6a3d9a}
\definecolor{jcbrown}{HTML}{b15928}
\definecolor{jclightred}{HTML}{fb8072}
\definecolor{jclightgreen}{HTML}{b3de69}
\definecolor{jclightblue}{HTML}{80b1d3}
\definecolor{jclightorange}{HTML}{fdb462}
\definecolor{jclightpurple}{HTML}{bebada}
\definecolor{jcredl}{HTML}{fb8072}
\definecolor{jcgreenl}{HTML}{b3de69}
\definecolor{jcbluel}{HTML}{80b1d3}
\definecolor{jcbluell}{HTML}{ece7f2}
\definecolor{jcorangel}{HTML}{fdb462}
\definecolor{jcpurplel}{HTML}{bebada}
\definecolor{jcbluem}{HTML}{488bb8}
\definecolor{jcyellow}{HTML}{ffff99}

\lstdefinestyle{mystyle}{
  frame=tblr,
  commentstyle=\color{codegreen},
  keywordstyle=\color{codepurple},
  basicstyle=\footnotesize\ttfamily,
  breakatwhitespace=false,
  breaklines=true,
  captionpos=b,
  keepspaces=true,
  numbers=left,
  numbersep=5pt,
  showspaces=false,
  showstringspaces=false,
  showtabs=false,
  tabsize=2,
  escapeinside={(*@}{@*)},
}
\lstdefinelanguage{CSL}{%
  language     = C,
  morekeywords = {var, param, i16, f32, fn, color, direction},
}

\lstdefinestyle{csl}{
  style=mystyle,
  basicstyle=\ttfamily\scriptsize,
  keywordstyle=\ttfamily\scriptsize,
  commentstyle=\color{black!45},
  numbers=none,
  frame=single,
  rulecolor=\color{black!30},
  framesep=3pt,
  breaklines=false,
  linewidth=\linewidth,
  aboveskip=0pt,
  belowskip=2pt,
}

\DeclareRobustCommand{\peshape}[1]{%
  \begin{tikzpicture}[x=2pt, y=2pt, baseline=0pt]
    \foreach \c/\r in {#1}
      {\fill[jcbluem, fill opacity=0.8] (\c,\r) rectangle ++(0.8,0.8);}
  \end{tikzpicture}%
}
\newcommand{\shaperect}{\peshape{0/0,1/0,2/0,0/1,1/1,2/1,0/2,1/2,2/2}}
\newcommand{\shapetetris}{\peshape{0/0,1/0,2/0,0/1,1/1,2/1,2/2,1/3,2/3}}

\usepackage{pifont}
\newcommand{\name}{WaferTetris\xspace}

\setcopyright{none}

\begin{document}

\title{WaferTetris: Efficient Freeform Mapping on Wafer-Scale Accelerators}



\author{Dayou Du}
\affiliation{%
  \institution{University of Edinburgh}
  \country{}
}
\email{dayou.du@ed.ac.uk}

\author{Congjie He}
\affiliation{%
  \institution{University of Edinburgh}
  \country{}
}
\email{congjie.he@ed.ac.uk}

\author{Andrew Boutros}
\affiliation{%
  \institution{University of Waterloo}
  \country{}
}
\email{andrew.boutros@uwaterloo.ca}

\author{Luo Mai}
\affiliation{%
  \institution{University of Edinburgh}
  \country{}
}
\email{luo.mai@ed.ac.uk}

\author{Vaughn Betz}
\affiliation{%
  \institution{University of Toronto}
  \country{}
}
\email{vaughn@eecg.utoronto.ca}

\author{Jianyi Cheng}
\affiliation{%
  \institution{University of Edinburgh}
  \country{}
}
\email{jianyi.cheng@ed.ac.uk}


\begin{abstract}
Wafer-scale accelerators offer massive compute capacity through
hundreds of thousands of processing elements (PEs) interconnected
across a single wafer, but realizing this potential requires
efficient application mapping.
Their scale makes programming these systems and mapping applications
onto them challenging.
Existing approaches rely on template-based mappings to maintain
productivity, but these templates restrict the mapping design space.
In particular, they impose fixed geometric layouts over the PE grid,
often leaving compute and memory resources underutilized.

In this work, we present \name, a system that removes these geometric
restrictions through freeform mapping.
Its Tetris-like strategy flexibly packs computation across the wafer,
improving PE utilization and usable memory capacity while preserving
PE-task code.
We develop a recurrence-based throughput model that captures branch-and-join
dependencies to guide spatial parallelism, and an architecture model with
per-color gates that enable existing FPGA place-and-route algorithms to
enforce WSE routing constraints.

We evaluate \name on Cerebras WSE-3 using LLM decoding and scientific
stencil benchmarks.
For LLM decoding, it achieves up to $4.42\times$ the system throughput of
a template-based automatic mapping baseline on the same fabric size.
Compared with the manually optimized WaferLLM, \name supports $1.8\times$
the context length for LLaMA3.1-8B (80k versus 44k), while delivering
equal or higher per-user throughput at matched context lengths,
up to $1.10\times$.
\end{abstract}

\maketitle

\section{Introduction}

\begin{figure}[!t]
\begin{subfigure}[b]{\linewidth}
\begin{subfigure}[b]{\textwidth}
\includegraphics[width=\textwidth]{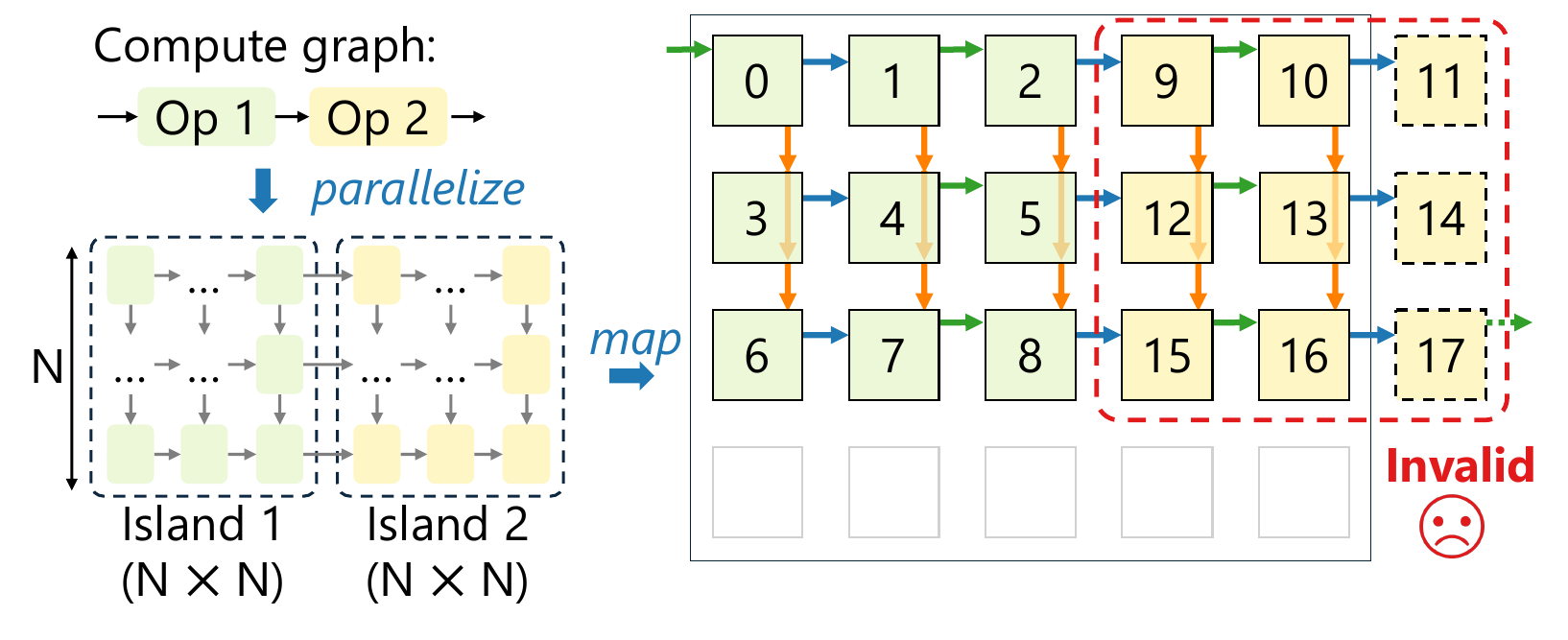}
\caption{
An example of mapping a computing task to a WSE for acceleration.
Traditional approaches are template-based and constrained by the PE island shapes, leading to spatial fragmentation.}
\label{fig:moti:invalid}
\end{subfigure}

\begin{subfigure}[b]{0.48\textwidth}
\centering
\includegraphics[width=\textwidth]{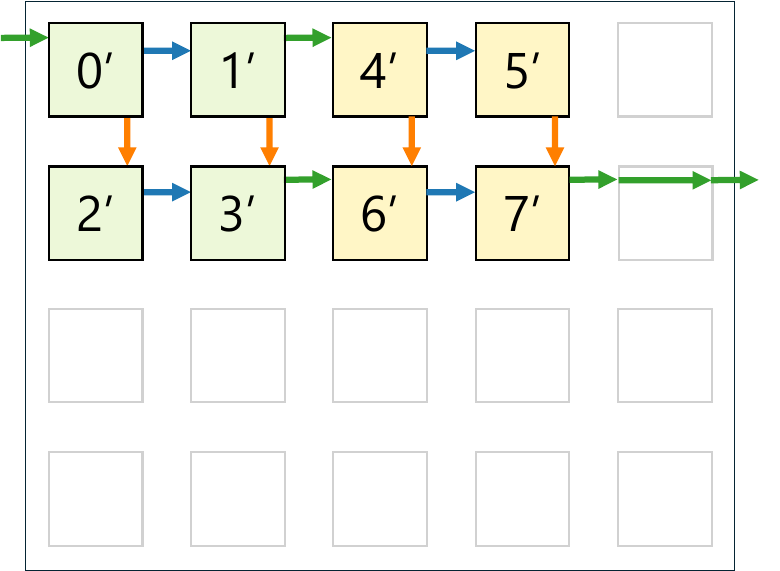}
\caption{Existing systems downscale PE island sizes to meet the geometric constraints, leading to limited spatial parallelism.}
\label{fig:moti:scaled}
\end{subfigure}
\hfill
\begin{subfigure}[b]{0.48\textwidth}
\centering
\includegraphics[width=\textwidth]{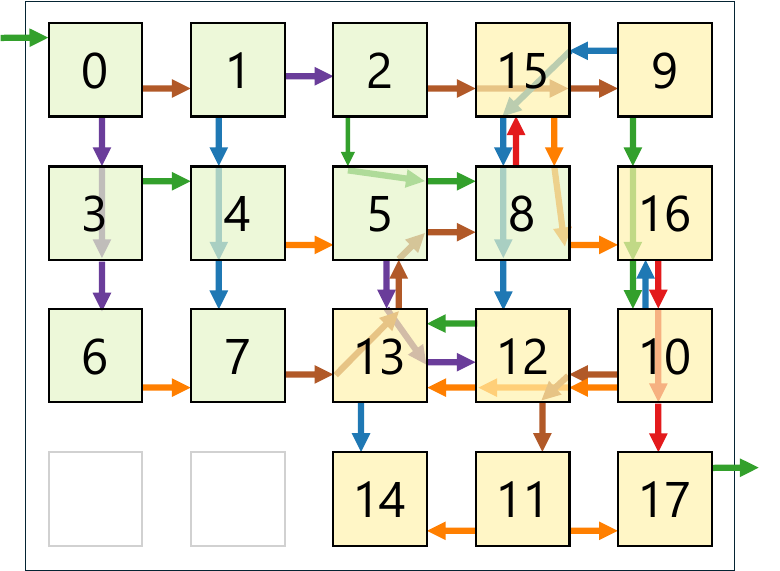}
\caption{Our work removes this restriction by enabling automated Tetris-like mapping, leading to improved spatial parallelism.}
\label{fig:moti:ourwork}
\end{subfigure}
\begin{subfigure}[b]{\textwidth}
\centering
\includegraphics[width=\linewidth]{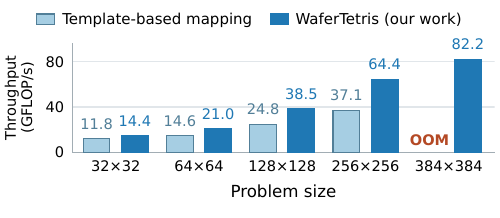}

\caption{Enabling Tetris-like mapping leads to significant throughput improvement, scaling with the workloads. OOM = out-of-memory.}
\label{fig:moti:results}
\end{subfigure}
\end{subfigure}
\caption{We present \name, an automated mapping system that improves resource utilization on a wafer-scale accelerator.}
\label{fig:moti}
\end{figure}

Wafer-scale accelerators have demonstrated substantial performance gains across a range of workloads, including large language model (LLM) training~\cite{dey2023cerebras}, LLM inference~\cite{he2025waferllm}, and scientific computing~\cite{stawinoga2026mlir}.
Emerging systems, such as the Cerebras Wafer-Scale Engine (WSE), integrate hundreds of thousands of processing elements (PEs) on a single silicon substrate, offering unprecedented compute density and low-latency on-chip communication.

Efficient spatial mapping is key to realizing this potential.
Mapping determines how many PEs execute each mapping unit (an individual operator or a fused kernel), where those PEs are placed on the fabric, and how data travels between them---decisions we refer to as spatial parallelism, placement, and routing.
Each PE has its own local memory, and communication uses finite routing resources along the mesh.
Consequently, mapping determines both the compute and memory capacity available to an application and the cost of exchanging data.
An efficient mapping must make effective use of these resources to maximize application throughput while respecting local memory and routing constraints~\cite{guo2023tapa,chen2024allo}.

Existing mapping approaches rely on templates whose geometric constraints can cause {\em spatial fragmentation}: free PEs remain available, but their spatial arrangement cannot accommodate the required island shapes.
Both automatic mapping systems, such as CU.POKer~\cite{jiang2020cu}, and manually optimized systems, such as WaferLLM~\cite{he2025waferllm}, instantiate mapping units as parameterized templates over rectangular PE grids, known as PE islands.
These templates save development effort by reusing computation and communication patterns, but couple them to restricted physical shapes.
For example, in \Cref{fig:moti:invalid}, island 2 fails to be mapped due to mismatches between its shape and the spatial layout, even though sufficient PEs are available at the bottom row of the WSE.
To obtain a feasible layout while preserving the templates, existing mapping systems scale down the islands~\cite{jiang2020cu, stawinoga2026mlir}, as illustrated in \Cref{fig:moti:scaled}, leaving many PEs unused.

These geometric restrictions limit compute parallelism, reduce usable memory capacity, and increase layout adaptation effort.
Shrinking islands leaves compute and local memory resources unused, limiting throughput and the workload sizes that fit on one WSE (\Cref{fig:moti:results}).
For example, on WSE-3, WaferLLM leaves approximately 344k of the 862.5k PEs in the usable fabric budget unallocated during LLaMA3.1-8B decoding, stranding about 40\% of the available peak floating-point compute capacity (FLOPS) and local SRAM capacity.
As a result, larger workloads may require multiple WSEs despite sufficient aggregate resources on one device.
Furthermore, changing a mapping unit's PE allocation also changes its island dimensions, potentially requiring new templates with rewritten PE-task code and customized communication paths.
This manual adaptation is labor-intensive and difficult to scale.

Our solution is to generate resource-efficient freeform mappings at practical compilation cost.
We define {\em freeform mapping} as placing and routing a logical PE-task graph without requiring each mapping unit to occupy a predefined geometric shape, while preserving the provided PE-task code and logical connectivity.
In \Cref{fig:moti:ourwork}, PEs with the same index perform identical tasks and exchange data with the same peers as in \Cref{fig:moti:invalid}; only their physical locations and communication routes change.
This flexibility allows island 2 to use the otherwise unused PEs along the bottom row without reducing its PE allocation.
Compared with template-based mapping, it increases throughput from 37.1 to 64.4\,GFLOP/s ($1.74\times$) for the $256\times256$ case.
Using 18 PEs instead of 8 also provides $2.25\times$ the aggregate local SRAM capacity, enabling the $384\times384$ case, where the template-based mapping runs out of memory (\Cref{fig:moti:results}).
However, freeform layouts can introduce longer communication paths and concentrate routing-resource demand at intermediate PE routers, requiring the mapper to navigate an enormous search space of coupled placement and routing decisions while maintaining communication locality and respecting routing capacity~\cite{he2025waferllm}.

To enable freeform mapping, our key insight is that mature FPGA place-and-route (P\&R) algorithms~\cite{betz1997vpr,elgammal2025vtr} are designed to meet these requirements.
FPGA placement reduces wirelength by keeping communicating blocks close, which limits the number of mesh hops and improves communication locality on the WSE.
FPGA routing resolves conflicts over finite routing resources; with an appropriate representation of WSE router resources, it can enforce capacity constraints at every PE router.
These efficient P\&R algorithms enable us to generate and compile freeform mappings within tens of minutes per configuration.

\begin{figure*}[t]
\centering
\begin{subfigure}[t]{0.28\textwidth}
\begin{lstlisting}[language=CSL, style=csl]
// pe.csl
var A: [M_local * N_local]f32;
var x: [N_local]f32;
fn op() void {
  x = receive_x();
  y = compute(A, x); // y = A*x
  send_y(y);
}\end{lstlisting}
\caption{The PE task of a distributed GEMV. Its workload sizes are parameters, so one task serves any island size.}
\label{fig:csl:pe_code}
\end{subfigure}
\hfill
\begin{subfigure}[t]{0.345\textwidth}
\begin{lstlisting}[language=CSL, style=csl]
layout {
  for (@range(p)) |r| {
   for (@range(p)) |c| {
    @set_tile_code(c, r, "pe.csl", .{
      .M_local=M/p, .N_local=N/p });
    @set_color_config(c, r, color1, ...);
  }}
}\end{lstlisting}
\caption{\shaperect\ Template-based layout for \Cref{fig:moti:scaled}. The {\tt for} bounds are the extent of the island, so its shape is always a rectangle and only \texttt{p} can vary.}
\label{fig:csl:template_island_code}
\end{subfigure}
\hfill
\begin{subfigure}[t]{0.345\textwidth}
\begin{lstlisting}[language=CSL, style=csl]
layout {
  @set_tile_code(0, 0, "pe.csl", island0);
  @set_tile_code(1, 0, "pe.csl", island0);
  ...
  @set_tile_code(3, 1, "pe.csl", island1);
  @set_color_config(0, 0, color1, ...);
  ...
}\end{lstlisting}
\caption{\shapetetris\ Customized layout for \Cref{fig:moti:ourwork}, generated by \name. One call per PE and per route expresses any island shape.}
\label{fig:csl:tetris_code}
\end{subfigure}
\caption{
CSL code for the mappings in \Cref{fig:moti:scaled,fig:moti:ourwork}.
A layout decides where each task runs and how data moves between PEs; the PE task in (a) is the same under both layouts. In (c), \texttt{island0} and \texttt{island1} specify the PE-task parameters for each island, including \texttt{M\_local} and \texttt{N\_local}.
}
\label{fig:csl}
\end{figure*}

However, building an automated mapping flow on top of these algorithms presents two challenges.
First, island sizes across mapping units must be chosen jointly, yet the cumulative cost of P\&R and compilation scales with the number of candidates, potentially extending the search to days.
Existing performance proxies based on the slowest unit~\cite{sohrabizadeh2022autodse} mispredict throughput when the graph forks and rejoins, leading to suboptimal island-size choices that limit end-to-end performance.
Second, conventional FPGA routing abstractions do not faithfully model WSE routers.
FPGA routes may bypass logic blocks and access interchangeable block pins, whereas every flow entering, leaving, or passing through a WSE PE must traverse its router and consume one of its limited colors.
Directly reusing the FPGA model can therefore yield invalid mappings whose routes cannot be implemented on the WSE.

To address these challenges, we design and implement \name, an automated system
that produces Tetris-like mappings for WSEs.
Given an application and its PE-task implementations, \name selects spatial
parallelism, placement, and routing, and emits a CSL
layout~\cite{selig2022cerebras_sdk} without altering the provided PE-task code.
Our contributions are threefold:
\begin{itemize}
    \item A recurrence-based performance model that predicts steady-state
    pipeline throughput before physical mapping. \name turns an application
    into a timed graph whose nodes carry mapping-unit latencies and whose
    edges carry stream costs, and takes its maximum cycle mean as the
    initiation interval, so branch-and-join dependences set the predicted
    throughput.

    \item A WSE architecture model that enables existing FPGA
    place-and-route algorithms to generate WSE-feasible mappings.
    The model maps WSE colors to routing tracks and adds a color-gate bank
    at each PE, with one unit-capacity gate per color.
    All flows entering, leaving, or passing through a PE on the same color
    must traverse that gate, exposing conflicts to the router as ordinary
    congestion.

    \item An evaluation on Cerebras WSE-3 covering LLM decoding and
    scientific stencil workloads. For LLM decoding, \name achieves
    $3.24$--$4.42\times$ the system throughput of the automatic mapping
    baseline on the same fabric size, with over 93\% PE utilization.
    Compared with WaferLLM, it supports $1.8\times$ the context length
    for LLaMA3.1-8B (80k versus 44k), with up to $1.10\times$ the
    per-user throughput at matched context lengths.
\end{itemize}

\section{Background and Motivation}

\subsection{WSE Architecture}
\label{sec:background:wse}

A WSE is a homogeneous two-dimensional grid of PEs on a single wafer.
Generations differ in scale, with WSE-3 integrating around 900k PEs in a 762$\times$1172 array~\cite{cerebras_chip_2026} and WSE-2 around 850k~\cite{he2025waferllm}, but they share the same topology.
Each PE contains a vector processor core, a local memory, and a router, as shown on the right of \Cref{fig:tooflow}.
The core executes an independent instruction stream, the local memory holds both its code and its data, and the router connects the core to the routers of the four adjacent PEs.
The architecture exposes no shared address space.
All data exchange between PEs is explicit message passing over the mesh network-on-chip (NoC) formed by the routers.
A message is forwarded hop by hop between adjacent routers, so memory access latency is highly non-uniform and scales with the number of hops separating the communicating PEs, spanning up to three orders of magnitude at wafer scale~\cite{he2025waferllm}.
Each message is tagged with a {\em color}, a virtual channel that determines both the route the message follows and the task it activates on arrival, and 21 colors are available to applications.
Finally, the fabric has no attached off-chip memory.
A PE holds 48\,kB of SRAM, of which roughly 40\,kB remains for data, so the aggregate memory available to an application is determined by the number of PEs it occupies.

\subsection{Application Mapping on the WSE}
Mapping an application onto the WSE comes down to three decisions, which we call spatial parallelism, placement, and routing: how many PEs execute each operation, where those PEs are located on the fabric, and which color and direction carry each data movement between them.
A valid mapping must fit each PE's code and data in local memory
and respect each router's color budget, while maximizing
steady-state pipeline throughput for dataflow
execution~\cite{guo2023tapa,chen2024allo}.
WSE is programmed in CSL, a C-like language, in which a program has two distinct parts, the {\em PE task} and the {\em layout}.
The PE task defines a PE's computation and local data.
For example, the distributed matrix-vector multiplication (GEMV)
task in \Cref{fig:csl:pe_code} receives {\tt x}, multiplies it by
the local matrix slice {\tt A}, and forwards the result.
Parameterized workload sizes let the same task support different
degrees of spatial parallelism.
The layout specifies task locations via {\tt set\_tile\_code}
and binds colors to routing directions at each PE via
{\tt set\_color\_config}.

Existing WSE mapping systems~\cite{stawinoga2026mlir, he2025waferllm} describe the {\em layout} with a parameterized template that instantiates the PE task on a rectangular grid of PEs, known as a PE island.
\Cref{fig:csl:template_island_code} is the template behind the mapping in \Cref{fig:moti:scaled}, where a nested {\tt for} loop places the GEMV task at every coordinate of a \texttt{p}$\times$\texttt{p} island and configures the routes between them.
The {\em layout} generated by \name for the custom mapping in \Cref{fig:moti:ourwork} is shown in \Cref{fig:csl:tetris_code}.
\name generates fully unrolled mapping descriptions, allowing any PE task to be placed on any physical PE, producing Tetris-like or more distributed layouts.
Each data movement path is also customized and instantiated individually, which removes the geometric constraints of traditional island templates and significantly improves performance.

\subsection{Spatial Fragmentation and Its Costs}
\label{sec:moti:fragmentation}

Template-based mapping requires each PE island to occupy a region
with a prescribed shape, typically a rectangle.
Even when enough PEs remain free in total, they may be scattered
across separate gaps or form an irregular region into which the
required island cannot fit.
We refer to this mismatch between available capacity and required
island geometry as \emph{spatial fragmentation}.
In \Cref{fig:moti:invalid}, this mismatch forces the mapper to shrink the islands, leaving many PEs unused in \Cref{fig:moti:scaled}.
This problem can arise across workloads composed of multiple PE islands and imposes three limitations.

\noindent\textbf{Limited compute parallelism.}
Reducing island sizes assigns fewer PEs to each mapping unit, leaving less hardware available to execute its work concurrently.
Consequently, an application can remain limited by compute parallelism even when PEs elsewhere on the fabric are idle.
\Cref{fig:moti:results} shows the performance impact: for the $256\times256$ case, the template-based mapping achieves 37.1\,GFLOP/s, compared with 64.4\,GFLOP/s for the freeform layout.

\noindent\textbf{Reduced usable memory capacity.}
Each PE contributes local SRAM, so unused PEs also leave memory capacity unavailable to the mapping.
Shrinking an island distributes its workload across fewer PEs, increasing per-PE memory pressure.
A workload can therefore exceed local memory limits even when the wafer has sufficient aggregate SRAM.
In \Cref{fig:moti:results}, the template-based mapping runs out of memory for the $384\times384$ case, while the freeform layout accommodates it.

\noindent\textbf{Increased layout adaptation effort.}
Changing a mapping unit's PE allocation also changes its island dimensions.
Each candidate allocation therefore requires finding a placement in which all islands fit together; satisfying the total PE budget alone is insufficient.
When the shapes do not fit, the mapper must revise the allocations, potentially sacrificing parallelism or memory capacity.
Exploiting the remaining space through new templates instead requires programmers to rewrite PE-task code and tailor communication paths.
Repeating this process across workloads and fabric configurations makes layout adaptation labor-intensive and difficult to scale.




\section{Key Insight and Challenges}
\label{sec:observations}

These limitations motivate freeform mapping that preserves PE-task
implementations and logical connectivity while allowing physical placements
and routes to adapt to the available fabric. Realizing this flexibility at wafer scale requires an efficient way to explore
PE placements and communication routes.
We identify an opportunity to adapt mature FPGA P\&R algorithms to this
task, together with two challenges that a mapping flow built on them
must still overcome.

\noindent\textbf{Insight: FPGA P\&R Objectives Align with WSE Communication Modeling.}
FPGA place-and-route (P\&R) takes a netlist describing logic blocks and
their connections, together with an architecture model of the target device.
The architecture model specifies the available placement sites and
represents routing resources as a graph with connectivity and capacity
constraints.

On the WSE, PEs exchange data through a mesh of routers connecting
their local memories.
WaferLLM highlights both the variation in communication latency
with physical distance and the limited routing resources available
at each PE~\cite{he2025waferllm}.
Effective WSE mapping must therefore account for both
communication locality and routing capacity.

FPGA P\&R optimizes closely related signals. Simulated-annealing placement keeps
communicating blocks close to reduce estimated wirelength and, in timing-driven
settings, path delay~\cite{betz1997vpr,elgammal2025vtr}. On the WSE's unit-hop
mesh, shorter wirelength corresponds to fewer hops and thus lower communication
latency. Negotiated-congestion routing iteratively resolves contention for
physical tracks. With tracks representing WSE colors and a model that captures
per-router resource use, this process can also resolve WSE color contention.

This alignment enables \name to reuse mature FPGA P\&R algorithms
for freeform mappings that use PEs and local memory left idle
by rectangular templates.
Preserving communication locality lets these mappings match
regular mappings' performance while relaxing geometric constraints
(\Cref{sec:exp:pr-effectiveness}).
Removing predefined island shapes also decouples spatial-parallelism
exploration from rectangle packing:
\name efficiently explores PE allocations under the wafer's resource
budget, deferring physical placement and routing to P\&R
(\Cref{sec:tile_scheduler}).

However, realizing these benefits requires addressing two challenges:


\definecolor{crit}{RGB}{192,57,43}

\tikzset{
  cyfig/.style = {font=\scriptsize},
  unit/.style   = {draw, rounded corners=1pt, minimum width=8mm,
                   minimum height=5.4mm, align=center, inner sep=1pt, line width=.4pt},
  hot/.style    = {unit, draw=crit, fill=crit!7, line width=.8pt},
  ed/.style     = {-{Latex[length=1.3mm]}, line width=.4pt},
  edfade/.style = {ed, draw=black!25},
  edhot/.style  = {ed, draw=crit, line width=.8pt},
  slab/.style   = {above, inner sep=1pt, font=\tiny},
  slabd/.style  = {below, inner sep=1pt, font=\tiny},
}

\begin{figure}[t]
\centering
\begin{subfigure}[b]{0.235\textwidth}
  \centering
  \begin{tikzpicture}[cyfig]
    \node[unit] (a1) at (0.00,0) {$U_1$\\[-2pt]\tiny $L\!=\!3$};
    \node[unit] (a2) at (1.00,0) {$U_2$\\[-2pt]\tiny $L\!=\!4$};
    \node[unit] (a3) at (2.00,0) {$U_3$\\[-2pt]\tiny $L\!=\!4$};
    \node[hot]  (a4) at (3.00,0) {$U_4$\\[-2pt]\tiny $L\!=\!6$};
    \draw[edfade] (a1) -- node[slab,black!30]{1} (a2);
    \draw[edfade] (a2) -- node[slab,black!30]{2} (a3);
    \draw[edfade] (a3) -- node[slab,black!30]{1} (a4);
    \draw[edfade] (a1) to[out=-48,in=-132] node[slabd,black!30]{$S\!=\!5$} (a3);
  \end{tikzpicture}
  \caption{slowest-unit}
  \label{fig:cyc:slowest}
\end{subfigure}
\hfill
\begin{subfigure}[b]{0.235\textwidth}
  \centering
  \begin{tikzpicture}[cyfig]
    \node[unit] (b1) at (0.00,0) {$U_1$\\[-2pt]\tiny $L\!=\!3$};
    \node[unit] (b2) at (1.00,0) {$U_2$\\[-2pt]\tiny $L\!=\!4$};
    \node[hot]  (b3) at (2.00,0) {$U_3$\\[-2pt]\tiny $L\!=\!4$};
    \node[unit] (b4) at (3.00,0) {$U_4$\\[-2pt]\tiny $L\!=\!6$};
    \draw[ed]    (b1) -- node[slab]{1} (b2);
    \draw[edhot] (b2) -- node[slab,crit]{2} (b3);
    \draw[edhot] (b3) -- node[slab,crit]{1} (b4);
    \draw[edhot] (b1) to[out=-48,in=-132] node[slabd,crit]{$S\!=\!5$} (b3);
  \end{tikzpicture}
  \caption{recurrence}
  \label{fig:cyc:ours}
\end{subfigure}
\caption{A slowest-unit objective reads only $L$ and reports $6$ at
$U_4$, while the actual interval of $12$ is modeled by the recurrence
graph.}
\label{fig:cycle-vs-slowest}
\end{figure}
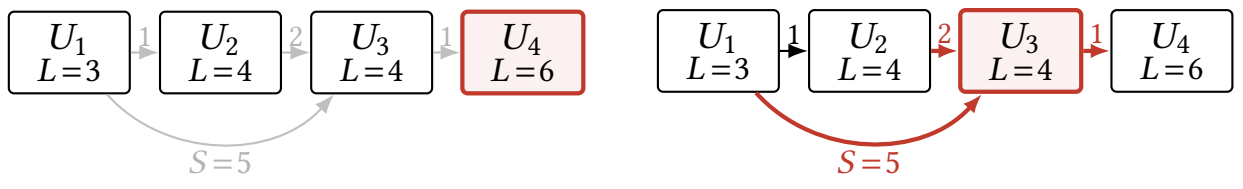

\begin{figure}[!t]
\centering
\begin{subfigure}[b]{0.235\textwidth}
  \centering
  \includegraphics[width=\linewidth]{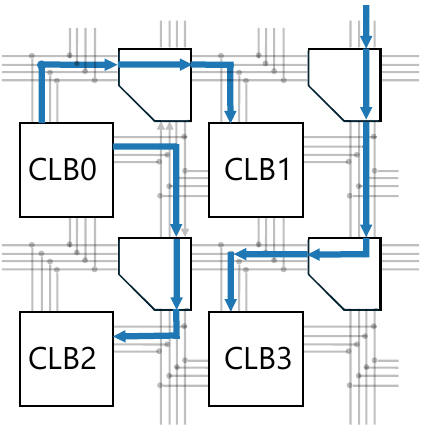}
  \caption{FPGA routing model}
  \label{fig:mismatch:fpga}
\end{subfigure}
\hfill
\begin{subfigure}[b]{0.235\textwidth}
  \centering
  \includegraphics[width=\linewidth]{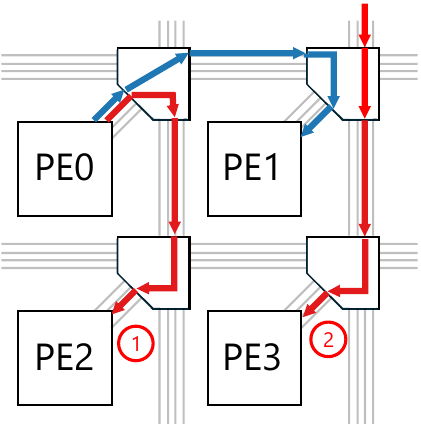}
  \caption{WSE routing model}
  \label{fig:mismatch:wse}
\end{subfigure}
\caption{Routing-resource mismatch between a conventional FPGA routing model and the WSE for the same logical communication. Markers 1 and 2 denote the I/O-access and router-semantics mismatches, respectively.}
\label{fig:mismatch}
\end{figure}

\begin{figure*}[t]
\centering
\includegraphics[width=\textwidth]{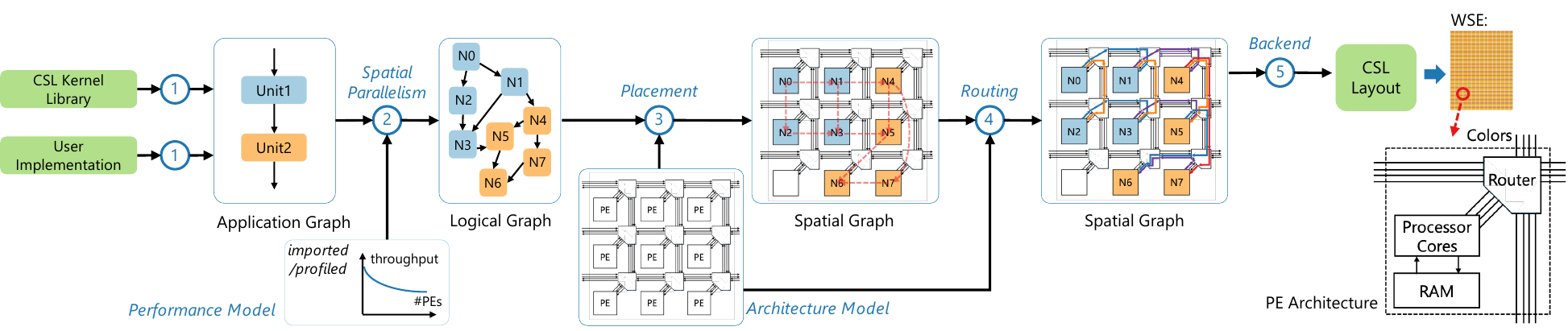}
\caption{Overview of the \name mapping workflow.}
\label{fig:tooflow}
\end{figure*}

\phantomsection
\label{chal:circular}
\noindent\textbf{Challenge 1: The Performance Estimation Gap.}
Given multiple mapping units under a shared wafer-scale PE budget, the island
size of each unit has to be chosen jointly with the others, so the number of
candidate assignments grows combinatorially with the application.
Scoring one candidate faithfully, however, means running the entire flow:
running P\&R, compiling the resulting CSL
program, and executing it on hardware or in simulation.
Each such flow may take hours, so measuring more than a handful of candidates
is out of reach.

Existing proxies do not close this gap. Spatial dataflow systems commonly balance
the pipeline by minimizing the slowest
unit~\cite{sohrabizadeh2022autodse,guo2023tapa,chen2024allo}, and CU.POKer~\cite{jiang2020cu}
likewise takes the maximum kernel execution time as the pipeline bottleneck.
This assumes the units run independently, which holds for a simple chain but
not for the branch-and-join structure common in modern applications, such as
the residual path of a transformer layer that forks one result to two units
and rejoins it later (\Cref{fig:cycle-vs-slowest}).
Such waiting propagates, so the interval is determined by a dependence path
spanning several units rather than by any single unit: in the figure the
objective predicts $6$ cycles while the pipeline completes one output every
$12$, and the unit it selects is not the one that limits throughput.

\noindent\textbf{Challenge 2: The Routing-Resource Modeling Gap.} Reusing FPGA P\&R algorithms requires an architecture model that
faithfully represents WSE routing constraints.
A route that is legal in the model must also be implementable on the hardware. However, conventional FPGA architecture models do not fully capture how traffic entering, leaving, or passing through a WSE PE consumes its limited router resources. Directly reusing these models can therefore admit routes that appear
legal but cannot be implemented on the WSE.
\Cref{fig:mismatch} illustrates two sources of this modeling gap. First, there is an \textbf{I/O access mismatch}:
when PE0 communicates with both PE1 and PE2, an FPGA-style CLB model may
realize the fanout using a single routing channel because the CLB exposes
multiple equivalent I/O pins, whereas a WSE PE has only one router-facing
access point and must serialize both flows through that interface.
Second, there is a \textbf{router-semantics mismatch}. FPGA routers commonly
represent wires and programmable switches as a routing-resource graph. In a
conventional island-style FPGA, these wires run beside the logic blocks, so a
signal may pass a CLB without consuming any routing resource inside it. On a
WSE, by contrast, every flow entering, leaving, or passing through a PE must
traverse the PE's router. Thus, if traffic to PE3 passes through PE1, it must
use a different color once PE1's router resource is occupied by another flow.
These mismatches lead to different resource accounting: the FPGA model need
not charge transit traffic to a per-block router, whereas the WSE mapping must
configure a color at every PE router along the path.

\section{Methodology}
\label{sec:method}



\subsection{\name Framework Overview}

\name maps an application graph
of user-defined mapping units and generates its CSL layout, as shown in
\Cref{fig:tooflow}. A mapping unit may represent a single operation, a fused
group of operations, or an entire model layer. Each unit is backed by either a
reusable implementation from the CSL kernel library or a user-provided kernel
implementation. \name preserves the provided PE-task code and customizes
its spatial allocation, placement, and communication routes.

\noindent
{\textcolor{jcblue}{\textcircled{\raisebox{-0.9pt}{1}}}}
The user constructs the application graph by selecting a CSL implementation
for each mapping unit and declaring its workload dimensions and data
dependencies. Each node denotes one mapping unit, while each edge denotes a
data dependency between units.

\noindent
{\textcolor{jcblue}{\textcircled{\raisebox{-0.9pt}{2}}}}
Given the application graph, \name uses a recurrence-based
throughput model to select spatial parallelism that maximizes
predicted end-to-end pipeline throughput
(\Cref{sec:tile_scheduler}).
The resulting logical graph represents individual PE tasks
as nodes and data movement between PEs as edges.

\noindent
{\textcolor{jcblue}{\textcircled{\raisebox{-0.9pt}{3}}}}-{\textcolor{jcblue}{\textcircled{\raisebox{-0.9pt}{4}}}}
\name models the WSE's interconnect topology and resource constraints
to enable mapping exploration with existing P\&R tools.
Given the logical graph and architecture model, the P\&R tool
produces a spatial graph specifying PE-task placements and
routes for inter-PE data movement.
We describe the architecture model in \Cref{sec:architecture_modeling}
and placement and routing in \Cref{sec:placement_and_routing,sec:routing}.

\noindent
{\textcolor{jcblue}{\textcircled{\raisebox{-0.9pt}{5}}}} 
The final spatial graph describes a valid mapping of PE tasks on the WSE.
Our back-end then performs a one-to-one translation of this spatial graph into an equivalent CSL layout program, working with the PE tasks from the input program to execute on the WSE.

We use the leading open-source P\&R tool, VTR 9.0~\cite{elgammal2025vtr}, for prototyping.
Both application graphs and logical graphs are represented in DOT format.
The architecture model is specified in an XML format following the existing representations in P\&R toolchains~\cite{betz1997vpr, shah2019yosys+}.
The spatial graph is also a DOT graph where the PE placement and routing constraints are encoded as metadata attached to the nodes and edges, respectively.

\subsection{Exploring Spatial Parallelism at Wafer Scale}
\label{sec:tile_scheduler}

In this section, we describe how \name explores spatial parallelism for
freeform mappings. Each mapping unit $i$ has a parallelism configuration
$\mathbf{s}_i=(h_i,w_i)$, denoting its logical PE-grid dimensions.
We first state the objective and the recurrences that bound
it (\Cref{sec:throughput_model}), then give the two weight models they are
built from (\Cref{sec:actor_comm_model}), and finally search the parallelism
space under the WSE's mapping constraints
(\Cref{sec:cp_solver}).

\subsubsection{Recurrence-Based Throughput Model}
\label{sec:throughput_model}

As \Cref{chal:circular} showed, balancing the pipeline by minimizing the
slowest mapping-unit
latency~\cite{sohrabizadeh2022autodse,guo2023tapa,chen2024allo} mispredicts the
achievable interval whenever the application graph forks and rejoins, because
the limit is then set by a dependence path spanning several units rather than
by any single unit.
\name therefore expresses spatial-parallelism exploration as
\begin{equation}
    \mathbf{s}^{*}
    = \operatorname*{arg\,min}_{\mathbf{s}\in\mathcal{S}}
    \widehat{\mathrm{II}}(\mathbf{s}),
    \qquad
    \mathbf{s}=\{\mathbf{s}_i\}_{i\in\mathcal{U}},
    \label{eq:parallelism_objective}
\end{equation}
where $\mathcal{U}$ is the set of mapping units, $\mathcal{S}$ is the set of
legal parallelism assignments, and $\widehat{\mathrm{II}}$ is the predicted
steady-state initiation interval---the fabric-cycle gap between consecutive
pipeline outputs after the pipeline is filled.

Given the dataflow nature of the WSE, we use dataflow analysis: the
initiation interval of a timed graph is its \emph{maximum cycle mean} (MCM),
the largest ratio of accumulated execution time to accumulated slack over the
directed recurrences of the
graph~\cite{reiter1968scheduling,karp1978characterization}. For the application
graph under an assignment $\mathbf{s}$, compute and transfer actions carry the
mapping-unit latency $L_i(\mathbf{s})$ and stream cost $S_e(\mathbf{s})$,
respectively:
\begin{equation}
\widehat{\mathrm{II}}(\mathbf{s})
= \max_{r\in\mathcal{R}(\mathbf{s})}
  \frac{\sum_{i\in\mathcal{A}(r)} L_i(\mathbf{s})
      + \sum_{e\in\mathcal{E}(r)} S_e(\mathbf{s})}
       {\sum_{e\in\mathcal{E}(r)} d_e},
\label{eq:pipeline_ii}
\end{equation}
Here $\mathcal{R}(\mathbf{s})$ is the set of directed recurrences under
assignment $\mathbf{s}$, and $\mathcal{A}(r)$ and $\mathcal{E}(r)$ are the
units and edges on recurrence $r$. To construct
$\mathcal{R}(\mathbf{s})$, \name expands each mapping unit into the ordered
receive--compute--send actions generated by the CSL back end, adds same-round
program and data dependences, and folds repeated executions by connecting each
buffer release to its next use. The slack $d_e$ is the maximum number of
complete rounds by which a producer may advance before the corresponding
storage is released, statically derived from buffer capacity, send/receive
order, and blocking or asynchronous communication
semantics~\cite{stuijk2006tradeoffs}; the resulting self-timed graph allows
independent actions to overlap, while join and capacity edges capture
synchronization and backpressure. In \Cref{fig:cycle-vs-slowest}, $U_3$
serially receives the 5- and 2-cycle streams, computes for 4 cycles, and sends
for 1 cycle before reusing its state, while $U_4$ overlaps with the next round,
yielding an initiation interval of $5+2+4+1=12$ cycles.

\subsubsection{Latency and Streaming Model}
\label{sec:actor_comm_model}

The two weight families in \Cref{eq:pipeline_ii} are modeled separately: a
latency model for $L_i$ of a unit on its island, and a
streaming model for the one-round cost $S_e$ of an edge between
two islands.

\noindent\textbf{Latency model $L_i(\mathbf{s})$.}
Motivated by the alignment between FPGA placement objectives and WSE
communication locality, \name approximates the latency of a freeform mapping
unit using its profile on a rectangle template. This avoids
running P\&R for every candidate during spatial-parallelism exploration; we
evaluate the approximation across two GEMV implementations and one GEMM
implementation in
\Cref{sec:exp:pr-effectiveness}.

For each mapping unit, let $(h,w)$ be its logical island dimensions and $W$ its
total compute workload. $C_h$ and $C_w$ denote the communication volumes along
the two island dimensions. \name models its local latency as
\begin{align}
L(h, w) = {}&{}\underbrace{\alpha_{\text{c}} \cdot \frac{W}{h \times w}
+ \alpha_{\text{m1}} \cdot \frac{C_h}{h}
+ \alpha_{\text{m2}} \cdot \frac{C_w}{w}}_{\text{workload}} \nonumber \\
&+ \underbrace{\alpha_{\text{d1}} \cdot h
+ \alpha_{\text{d2}} \cdot w}_{\text{data movement distance}} + \alpha_0,
\label{eq:latency_template}
\end{align}
where the $\alpha$ terms are fitted coefficients for the implementation and
target WSE. The workload terms capture compute work and communication volume
per partition, while the distance terms capture the logical span of
intra-island collectives. Here, $h$ and $w$ describe the logical template
rather than physical distances after P\&R. The full form applies to
two-dimensional kernels such as GEMM and GEMV; one-dimensional reductions such
as Softmax and RMSNorm, and neighbor-only kernels such as Stencil, use
corresponding specializations.

\name obtains these coefficients through an offline profiling process. For each
CSL implementation, it instantiates regular templates over representative
workloads and legal $(h,w)$ pairs, executes them on the target WSE, and fits
the coefficients in \Cref{eq:latency_template}; each measurement includes local
computation, intra-island communication, and synchronization. The fitted models
are stored with CSL kernel implementations and reused across applications.
\name provides models for common library kernels, including GEMM, GEMV,
Softmax, RMSNorm, and Stencil; user-provided kernels follow the same profiling
interface. Once fitted, evaluating a candidate configuration requires only a
model query.

\noindent\textbf{Streaming model $S_e(\mathbf{s})$.}
A logical edge $e$ moves one tensor between two islands that may cut it
into different numbers of slices. The overlap of those slices gives, for
each participating PE $p$, how many counterparts it talks to
($n_{e,p}$) and how many elements it streams ($V_{e,p}$). At one element
per cycle,
\begin{equation}
S_e(\mathbf{s})
= \max_{p\in P_e}
  \bigl(\alpha_{\mathrm{s}}\, n_{e,p}(\mathbf{s}) + V_{e,p}(\mathbf{s})\bigr),
\label{eq:streaming_cost}
\end{equation}
where $P_e$ is the set of sending and receiving PEs on $e$, and
$\alpha_{\mathrm{s}}$ is a per-connection setup cost from profiling.
The max, not the total volume, is what matters: a PE that overlaps
several counterparts pays several setups.

\subsubsection{Parallelism Exploration with Recurrence-Guided Refinement}
\label{sec:cp_solver}

To make the exploration tractable, \name combines a mixed-integer linear
programming (MILP) initialization with recurrence-guided refinement. Each mapping unit
$i$ chooses one supported parallelism configuration from its legal set $S_i$:
\begin{align}
    \forall i\in\mathcal{U}\ldotp (h_i,w_i)\in S_i.
\end{align}
The assignment must fit both the WSE fabric and the local SRAM of every PE:
\begin{align}
    \sum_{i\in\mathcal{U}} h_iw_i \leq F,\qquad
    \forall i\in\mathcal{U}\ldotp m_i(h_i,w_i)\leq\hat{m},
\end{align}
where $F$ is the available PE count and $\hat{m}$ is the per-PE SRAM capacity.
Communication-compatible island dimensions are enforced between connected
units.

Because the active recurrences in \Cref{eq:pipeline_ii} depend on the
parallelism configuration, the MILP provides an initial feasible assignment.
We define $T_{\mathrm{proxy}}\geq L_i$ for every unit and solve
\begin{equation}
\operatorname*{lex\,min}
\left(T_{\mathrm{proxy}},-\sum_i h_iw_i\right),
\label{eq:parallelism_proxy}
\end{equation}
first minimizing the proxy interval and then maximizing PE use. 

Starting from this assignment, recurrence-guided refinement enumerates
alternative dimensions for each matching-connected group and alternative
$(h,w)$ pairs for each mapping unit. For every candidate, \name recomputes $L_i$,
$S_e$, and the active recurrences in \Cref{eq:pipeline_ii}, using the
CSL-derived recurrence topology and slack $d_e$. Candidates that violate
hardware resource constraints are discarded.

Candidates are ranked lexicographically by predicted II, PE count, and the sum
of recurrence bounds. Each pass accepts the best improving legal candidate and
stops when none exists or after a fixed iteration budget.

\begin{table}[t]
    \centering
    \captionof{table}{Correspondence between WSE resources and their FPGA abstractions in our architecture model.}
    \label{tab:architecture_comparison}
{\footnotesize
\begin{tabular}{lcc}
\toprule
\makecell[l]{\bf Compute} & \bf FPGAs & \bf WSEs \\
\midrule
Component scales & \textasciitilde10$^6$ CLBs & \textasciitilde10$^6$ PEs \\
Multi-in multi-out functions & Yes & Yes \\
\midrule
\makecell[l]{\bf Communication} & \bf FPGAs & \bf WSEs \\
\midrule
Interconnect nodes & Switchbox & Router \\
Data channels & Hardware tracks & Virtual colors \\
Distributing data mechanisms & Fan-out & Broadcast  \\
Latency of the route & Wire length & Hop count \\
I/O pads & Edges or columns & Edges  \\
\bottomrule
\end{tabular}}
\end{table}

\subsection{Architecture Model for WSE-specific Mapping}
\label{sec:architecture_modeling}

In this section, we describe how \name adapts mature FPGA P\&R techniques to WSE mapping.
First, we model the WSE as an FPGA variant through a P\&R-compatible architecture
model that captures its PE array. We then align
placement for WSE communication locality and introduce WSE-specific routing
constraints to address the router-behavior mismatch identified in Challenge~2.
Finally, we leverage floorplanning to decompose wafer-scale P\&R into tractable
subproblems and improve mapping scalability.

\subsubsection{Modeling the WSE as an FPGA Variant}

To reuse mature resource-constrained FPGA P\&R techniques, \name models the WSE
as an FPGA variant through the correspondences summarized in
\Cref{tab:architecture_comparison}. At the compute level, each PE maps to a
custom CLB: both occupy one site in a two-dimensional spatial fabric and expose
a multi-input, multi-output interface. Our custom CLB uses a LUT as a
combinational crossbar, with application-derived ports and a path from every input to every
output. A subset switchbox preserves track identity across channel-to-channel
connections, but alone cannot prevent the pin connectivity at a PE from
forcing a route onto a different track. The crossbar LUT removes this
restriction by allowing any PE input to reach any PE output, so an end-to-end
route can remain on one indexed track and that track can consistently represent
one WSE color. We omit flip-flops because the application graph already treats
each PE as an atomic compute task; the P\&R stage places the whole task and
routes its external communications rather than synthesizing or packing the
PE's internal sequential logic.

At the communication level, each logical data stream becomes a bit-level net
that captures connectivity rather than payload values. A WSE router maps to a
switchbox, a routable color maps to a physical track, and WSE broadcast maps to
FPGA fan-out. Correspondingly, route hop count maps to wirelength. At the top
level, a parameterized fixed layout captures the target WSE fabric size: its
interior dimensions match the physical PE grid, while boundary tiles provide
I/O pads restricted to the left and right edges. Together, these
correspondences express WSE mapping as placing compute blocks on a fixed grid
and assigning their communication to a bounded set of routing resources.


\subsubsection{Aligning Placement for Communication Locality}
\label{sec:placement_and_routing}

Placement operates on individual PE tasks, with nets representing
both intra-unit and inter-unit communication. Thus, even with
PE allocations fixed, the placer optimizes communication locality within each mapping unit as well as between units.
A unit's physical shape emerges from this placement rather than
being prescribed as a rectangle.

Communication latency on the WSE grows with the number of router hops, making
hop distance a first-order placement objective~\cite{he2025waferllm,wang2026temp}.
FPGA placers therefore keep communicating blocks close to reduce
wirelength~\cite{elgammal2021rlplace,li2019elfplace}.
\name aligns this objective with an FPGA simulated-annealing placer by modeling
the wafer interconnect as a regular unit-hop mesh with uniform channels. Under
this abstraction, placement wirelength serves as a direct proxy for hop count:
moves that bring communicating PEs closer reduce both the placement cost and
the expected communication latency.

\begin{figure}[t]
\centering
\begin{subfigure}[t]{0.23\textwidth}
    \centering
    \includegraphics[scale=0.45]{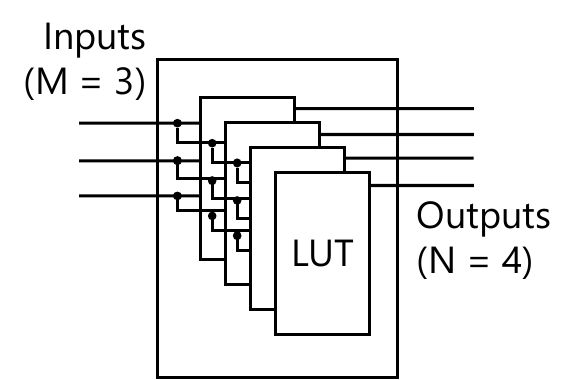}
    \caption{Modeling a WSE PE as a custom CLB with application-derived ports and a crossbar LUT.}
    \label{fig:arch:comp}
\end{subfigure}
\hfill
\begin{subfigure}[t]{0.23\textwidth}
    \centering
    \includegraphics[scale=0.45]{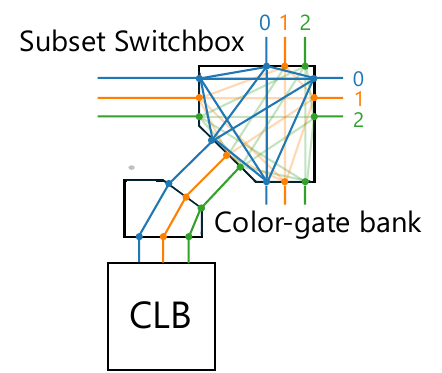}
    \caption{Connecting a modeled CLB to a subset switchbox through a per-color gate bank.}
    \label{fig:arch:router}
\end{subfigure}
\caption{WSE-specific architecture abstraction. \name represents PEs as custom CLBs and router colors as subset-switch tracks guarded by per-color gates, exposing WSE compute and routing constraints to existing P\&R tools.}
\label{fig:arch}
\end{figure}

\subsubsection{Aligning Routing with Resource Constraints}
\label{sec:routing}

Our routing abstraction aims to expose WSE color feasibility directly to
conventional FPGA routing algorithms. \name therefore binds routing tracks to
WSE colors one-to-one: track $c$ represents color $c$, and the channel width
equals the number of available colors. Uniform subset switchboxes preserve the
track index across hops and turns. Consequently, assigning a communication
path to track $c$ simultaneously assigns its WSE route to color $c$, allowing
track contention to represent color contention.

Track--color binding alone, however, does not eliminate the FPGA-specific
shortcuts identified in Challenge~2. To address the
I/O-access mismatch, \name models each PE as a custom block with
application-dependent input and output ports, all exposed on its router-facing
side (\Cref{fig:arch:comp}). These ports connect to uniform subset channels
through the PE router rather than through interchangeable tile sides. This
constrained interface preserves the communication degree of the logical graph
while excluding pin-access patterns unavailable on the WSE.

To address the router-semantics mismatch, \name augments every modeled
CLB with a \emph{color-gate bank} at its router interface, as shown in
\Cref{fig:arch:router}.
For each CLB $p$, which represents one WSE PE, and each color $c$,
the bank contains a unit-capacity gate $g_{p,c}$.
Each gate is an auxiliary routing-resource node shared by the PE's
local interface and the color-$c$ tracks in all four mesh directions.
\name removes direct bypasses around these gates, requiring
data produced by the PE, data consumed by the PE, and data forwarded
through the PE on color $c$ to traverse the same $g_{p,c}$.
Thus, distinct nets using different incident links still contend
for the same gate if they use the same color at the same PE.
The gate converts the WSE rule that distinct flows cannot independently
configure the same color at one PE into an ordinary unit-capacity
routing constraint.

This abstraction naturally matches negotiated-congestion routing
algorithms~\cite{betz1997vpr}. If multiple flows temporarily select the same
gate, its overuse increases the present and historical congestion costs.
Rip-up and reroute then moves a conflicting flow to another track---and hence
another color---or to an alternative path. The routing converges to a
WSE-feasible solution only when every color gate has an occupancy of at most
one, so the resulting resource assignment directly respects the PE-local
color constraints.

\begin{figure}[t]
\centering
\begin{subfigure}[b]{0.487\linewidth}
  \centering
  \includegraphics[width=\linewidth]{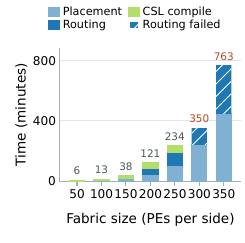}
  \caption{Scalability of P\&R}
  \label{fig:scalability}
\end{subfigure}
\hfill
\begin{subfigure}[b]{0.487\linewidth}
  \centering
  \includegraphics[width=\linewidth]{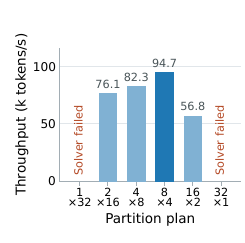}
  \caption{Effect of the partition plan}
  \label{fig:floorplanning}
\end{subfigure}
\caption{(a) The P\&R time scales exponentially with the fabric size. \name improves scalability via floorplanning. (b) Different floorplanning strategies could affect throughput results for LLaMA3.1-8B (sequence length = 4k). }
\label{fig:partition}
\end{figure}

\subsubsection{Floorplanning-Guided Mapping Exploration} \label{sec:floorplanning}

A known issue for P\&R optimization is scalability, as the mapping space grows combinatorially with millions of components in fabrics. This challenge directly carries over to \name, as illustrated in \Cref{fig:scalability}.
Prior FPGA floorplanning work~\cite{guo2021autobridge,guo2022rapidstream} shows that partitioning the fabric and performing independent P\&R could reduce compilation time while preserving high performance. 
Inspired by this, \name introduces a floorplanning interface that decomposes global P\&R into tractable subproblems and orchestrates the P\&R tool for scalable exploration.

Automated application partitioning remains an open problem~\cite{guo2022rapidstream}. 
By default, \name statically analyzes the application graph and partitions it into subgraphs balanced by PE count.
For fabric partitioning, it explores predefined templates and selects the best configuration using the latency model in \Cref{sec:tile_scheduler}.
\Cref{fig:floorplanning} shows this process on LLaMA3.1-8B (32 layers) with a sequence length of 4K.
A template $R \times C$ tiles the wafer into an $R$-by-$C$ grid of rectangular partitions: $1 \times 32$ is 32 vertical strips, and $32 \times 1$ is 32 horizontal strips.
Both of those extremes fail to yield valid mappings, while $8 \times 4$ achieves the best performance.
In our experiments, we therefore partition the fabric into an $8 \times 4$ grid for P\&R.

\name calls the P\&R tool sequentially on each subgraph–partition pair.
The resulting placements are then composed and globally routed to produce a complete mapping.

\FloatBarrier
\section{Experiments}

We compare the application throughput of \name and template-based mappings
under the same fabric size. We run experiments on
commercial Cerebras WSE-3 hardware. We also report PE
utilization, defined as the fraction of available PEs allocated to the
application, to quantify the use of available spatial parallelism.

Existing automatic mapping systems either target a different optimization
scope or lack publicly available implementations. The MLIR
pipeline~\cite{stawinoga2026mlir} targets single-operation optimization, while
we focus on system-level mapping. RAM-Wafer~\cite{dai2025ram} is specialized
for AI training rather than the decoding and stencil workloads evaluated
here. CU.POKer~\cite{jiang2020cu} and GigaPlacer~\cite{li2021placement} are
closed-source. We therefore reimplement the published template-based
algorithm from CU.POKer, the stronger of the two~\cite{ispd2021contest}, and
complete its inter-island routing with Cerebras' official
APIs~\cite{cerebras_wse3_libraries_2025} to obtain an executable baseline.

We first show that \name achieves up to $4.42\times$ the system
throughput of CU.POKer and increases PE utilization by up to 79.8\%
(\Cref{sec:exp:main_results}). Against the manually optimized
WaferLLM system~\cite{he2025waferllm}, \name achieves up to $1.10\times$
the per-user throughput and increases PE utilization by up to 50.1\%
(\Cref{sec:exp:waferllm}).

We then analyze the sources of performance gains and the cost of obtaining
a mapping through three questions:
\begin{itemize}
\item How effectively does FPGA P\&R preserve mapping-unit latency under
freeform placement? (\Cref{sec:exp:pr-effectiveness})
\item How accurate are the performance predictions, and how do
recurrence-guided optimization and freeform mapping contribute to
throughput gains?
(\Cref{sec:exp:analysis})
\item How much time is spent on profiling, scheduling, placement, routing,
and compilation?
(\Cref{sec:exp:mapping_time})
\end{itemize}

\subsection{Benchmarks}
\label{sec:exp:benchmarks}

We evaluate three benchmark groups covering LLM decoding and scientific
stencil computation. The two LLM groups use different mapping-unit
granularities: individual operators and fused kernels. Both evaluate
full-model decoding, with system throughput used for the CU.POKer
comparison and per-user throughput used for the WaferLLM comparison.

\noindent\textit{1) Operator-graph LLM decoding:}
We follow the graph-based mapping formulation of CU.POKer~\cite{jiang2020cu},
representing each decoder layer as a graph of operators mapped onto
separate PE islands. We evaluate Qwen2.5-7B, LLaMA3.1-8B, and LLaMA2-13B
at sequence lengths of 1k, 2k, and 4k, comparing \name with CU.POKer in
system decoding throughput, which is the total number of
output tokens generated per second across all concurrent decoding requests.
\Cref{fig:llama-layer-placement} shows a
representative LLaMA layer graph and its placements under CU.POKer and \name.

\definecolor{llamaInk}{HTML}{2F3740}
\definecolor{llamaUnused}{HTML}{F7F8FA}
\definecolor{llamaGrid}{HTML}{D9DEE3}
\definecolor{llamanorm}{HTML}{A9C5CF}
\definecolor{llamaq}{HTML}{CFC3D9}
\definecolor{llamak}{HTML}{6F5A8F}
\definecolor{llamav}{HTML}{E8C38C}
\definecolor{llamaqk}{HTML}{D9904F}
\definecolor{llamasoftmax}{HTML}{E2A9A6}
\definecolor{llamapv}{HTML}{A94F4F}
\definecolor{llamao}{HTML}{E7DFA8}
\definecolor{llamapost}{HTML}{86614F}
\definecolor{llamagate}{HTML}{C9DCB9}
\definecolor{llamaup}{HTML}{81B784}
\definecolor{llamadown}{HTML}{6E9FBE}

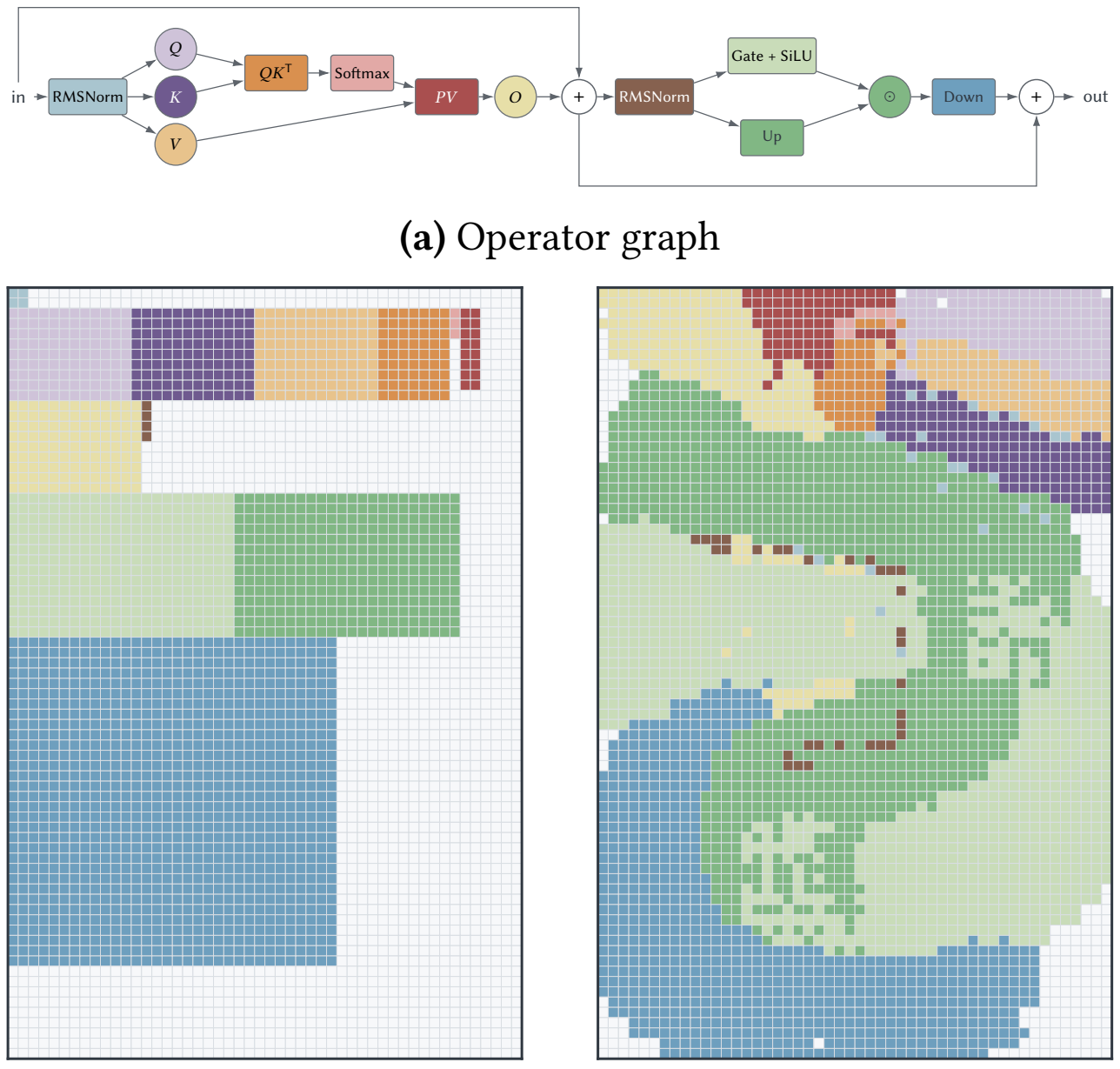
\begin{figure}[!t]
  \centering
  \begin{subfigure}[t]{\linewidth}
    \centering
    \resizebox{\linewidth}{!}{%
      \begin{tikzpicture}[
        x=1cm,
        y=1cm,
        font=\scriptsize\sffamily,
        flow/.style={-{Latex[length=1.7mm,width=1.1mm]},
                     line width=0.48pt, draw=llamaInk!78},
        op/.style={draw=llamaInk!72, rounded corners=1.2pt,
                  minimum width=9.5mm, minimum height=5.4mm,
                  inner sep=1.5pt, line width=0.48pt, align=center},
        mathop/.style={circle, draw=llamaInk!72, minimum size=6.3mm,
                      inner sep=0.6pt, line width=0.48pt},
        add/.style={circle, draw=llamaInk!78, fill=white,
                   minimum size=5.2mm, inner sep=0pt,
                   line width=0.55pt, font=\footnotesize},
        endpoint/.style={font=\footnotesize\sffamily, text=llamaInk}
      ]
        \node[endpoint] (xin) at (0,0) {in};
        \node[op,fill=llamanorm] (pre) at (1.05,0) {RMSNorm};
        \node[mathop,fill=llamaq] (q) at (2.38,0.72) {$Q$};
        \node[mathop,fill=llamak,text=white] (k) at (2.38,0) {$K$};
        \node[mathop,fill=llamav] (v) at (2.38,-0.72) {$V$};
        \node[op,fill=llamaqk] (score) at (3.90,0.36) {$QK^{\mathsf T}$};
        \node[op,fill=llamasoftmax] (softmax) at (5.20,0.36) {Softmax};
        \node[op,fill=llamapv,text=white] (mix) at (6.48,0) {$PV$};
        \node[mathop,fill=llamao] (oproj) at (7.52,0) {$O$};
        \node[add] (addattn) at (8.48,0) {$+$};
        \node[op,fill=llamapost,text=white] (post) at (9.62,0) {RMSNorm};
        \node[op,fill=llamagate] (gate) at (11.40,0.62) {Gate + SiLU};
        \node[op,fill=llamaup,text=llamaInk] (up) at (11.40,-0.62) {Up};
        \node[mathop,fill=llamaup,text=llamaInk] (mul) at (13.18,0) {$\odot$};
        \node[op,fill=llamadown,text=llamaInk] (down) at (14.30,0) {Down};
        \node[add] (addffn) at (15.40,0) {$+$};
        \node[endpoint] (yout) at (16.30,0) {out};

        \draw[flow] (xin) -- (pre);
        \draw[flow] (pre) -- (q);
        \draw[flow] (pre) -- (k);
        \draw[flow] (pre) -- (v);
        \draw[flow] (q) -- (score);
        \draw[flow] (k) -- (score);
        \draw[flow] (score) -- (softmax);
        \draw[flow] (softmax) -- (mix);
        \draw[flow] (v) -- (mix);
        \draw[flow] (mix) -- (oproj);
        \draw[flow] (oproj) -- (addattn);
        \draw[flow] (xin.north) -- (0,1.35) -- (8.48,1.35) --
                    (addattn.north);
        \draw[flow] (addattn) -- (post);
        \draw[flow] (post) -- (gate);
        \draw[flow] (post) -- (up);
        \draw[flow] (gate) -- (mul);
        \draw[flow] (up) -- (mul);
        \draw[flow] (mul) -- (down);
        \draw[flow] (down) -- (addffn);
        \draw[flow] (addffn) -- (yout);
        \draw[flow] (addattn.south) -- (8.48,-1.35) --
                    (15.40,-1.35) -- (addffn.south);
      \end{tikzpicture}%
    }
    \phantomsubcaption
    \caption*{\textbf{(a)} Operator graph}
    \label{fig:llama-layer:graph}
  \end{subfigure}

  \vspace{1mm}
  \begin{subfigure}[t]{0.47\linewidth}
    \centering
    \includegraphics[width=\linewidth]{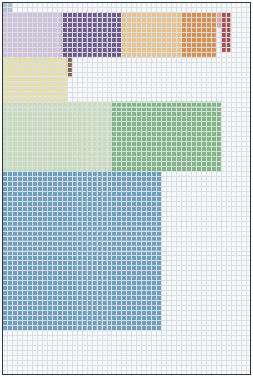}
    \phantomsubcaption
    \caption*{\textbf{(b)} CU.POKer}
    \label{fig:llama-layer:cupoker}
  \end{subfigure}
  \hfill
  \begin{subfigure}[t]{0.47\linewidth}
    \centering
    \includegraphics[width=\linewidth]{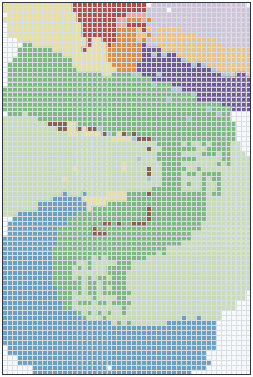}
    \phantomsubcaption
    \caption*{\textbf{(c)} \name}
    \label{fig:llama-layer:tetris}
  \end{subfigure}

  \caption{An example showing how CU.POKer and \name map a LLaMA layer
  onto a $50\times75$ PE region. Matching colors denote the same operators
  across the graph and both placements; white cells indicate unused PEs.}
  \Description{Panel (a) shows the branch-and-join operator graph of one
  LLaMA layer. Panels (b) and (c) show the corresponding CU.POKer
  rectangular-island placement and WaferTetris freeform placement on the
  same 50 by 75 PE region. Matching colors denote the same operators.}
  \label{fig:llama-layer-placement}
\end{figure}

\noindent\textit{2) Fused-kernel LLM decoding:}
We use WaferLLM's optimized fused kernels as mapping units and apply \name
to their spatial parallelism, placement, and routing. We report results for LLaMA3.1-8B and LLaMA2-13B, varying sequence length to examine decoding
performance and KV-cache capacity. We compare per-user decoding throughput
with the manually optimized WaferLLM implementation~\cite{he2025waferllm}.

\noindent\textit{3) Stencil computation:}
We evaluate two benchmarks spanning multiple PE islands:
Gray-Scott reaction-diffusion~\cite{pearson1993complex} and elastic wave
propagation~\cite{virieux1986psv}.
Gray-Scott is a two-stage pipeline combining parallel five-point
Laplacians with point-wise reactions, for four operators per timestep.
Elastic wave propagation uses a five-operator velocity--stress pipeline
with fourth-order finite-difference stencils~\cite{louboutin2019devito}.
We evaluate both benchmarks at Z-dimension sizes of 50, 100, and 200 and
compare throughput in lattice updates per second against CU.POKer.



\FloatBarrier
\subsection{Comparison against Template-based Approach}
\label{sec:exp:main_results}

\begin{figure}[!t]
  \centering
  \includegraphics[width=0.37345\linewidth]{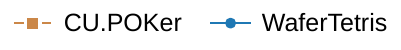}\par\smallskip
  \begin{subfigure}[b]{0.36723\linewidth}
    \centering
    \includegraphics[width=\linewidth]{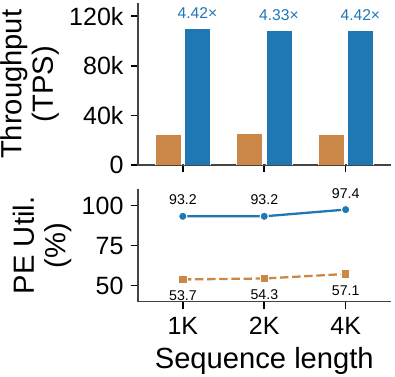}
    \caption{Qwen2.5-7B}
    \label{fig:overall:qwen}
  \end{subfigure}\hfill
  \begin{subfigure}[b]{0.30638\linewidth}
    \centering
    \includegraphics[width=\linewidth]{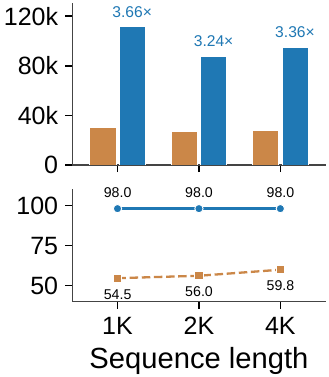}
    \caption{LLaMA3.1-8B}
    \label{fig:overall:llama8b}
  \end{subfigure}\hfill
  \begin{subfigure}[b]{0.30638\linewidth}
    \centering
    \includegraphics[width=\linewidth]{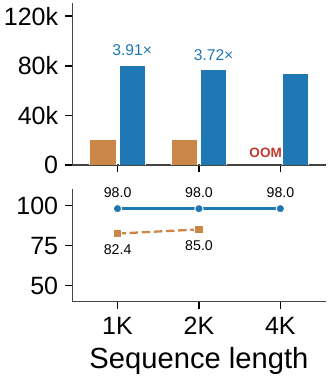}
    \caption{LLaMA2-13B}
    \label{fig:overall:llama13b}
  \end{subfigure}
  \caption{System throughput and PE utilization for LLM decoding on WSE-3.
  TPS = tokens/s.}
  \Description{Three model panels compare CU.POKer and WaferTetris for
  Qwen2.5-7B, LLaMA3.1-8B, and LLaMA2-13B at sequence lengths of 1K, 2K,
  and 4K. Each panel shows throughput bars above PE-utilization curves.
  WaferTetris achieves 3.24 to 4.42 times the throughput where both
  methods run; CU.POKer runs out of memory for LLaMA2-13B at 4K.}
  \label{fig:overall:llms}
\end{figure}

\begin{figure}[!t]
  \centering
  \includegraphics[width=0.39087\linewidth]{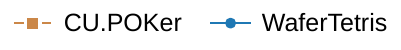}\par\smallskip
  \begin{subfigure}[b]{0.52184\linewidth}
    \centering
    \includegraphics[width=\linewidth]{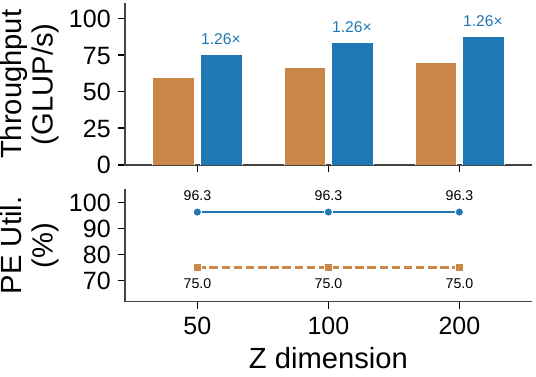}
    \caption{Gray--Scott}
    \label{fig:overall:gray}
  \end{subfigure}\hfill
  \begin{subfigure}[b]{0.45816\linewidth}
    \centering
    \includegraphics[width=\linewidth]{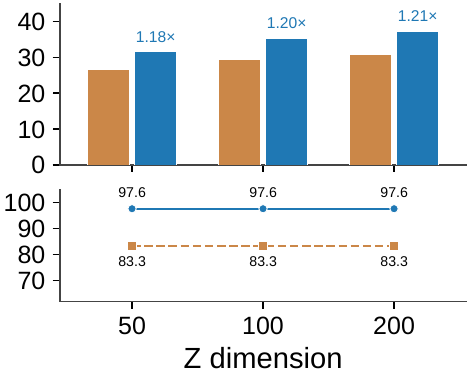}
    \caption{Elastic wave}
    \label{fig:overall:elastic}
  \end{subfigure}
  \caption{Stencil throughput and PE utilization on WSE-3.
  GLUP/s = billions of lattice updates per second.}
  \Description{Gray-Scott and elastic-wave panels compare CU.POKer with
  WaferTetris at Z dimensions of 50, 100, and 200. Throughput bars sit
  above PE-utilization curves. WaferTetris improves throughput by
  approximately 1.26 times for Gray-Scott and 1.18 to 1.21 times for
  elastic wave. PE utilization is constant across the Z dimensions.}
  \label{fig:overall:stencils}
\end{figure}

We compare \name with CU.POKer, an automatic mapping system based on templates,
on WSE-3 in \Cref{fig:overall:llms,fig:overall:stencils}.
\name uses a 98\% PE budget, reserving 2\% for routing resources.
Across the LLM
configurations where both methods run, \name achieves $3.24$--$4.42\times$
the system throughput of CU.POKer. These gains stem from two factors:
better island-size selection enabled by an accurate performance model,
and higher spatial parallelism enabled by freeform mapping.
\Cref{sec:exp:model_accuracy} analyzes these factors.

The WaferTetris mappings use 93.2--97.4\% of the available PEs for Qwen2.5-7B
and 98.0\% for both LLaMA3.1-8B and LLaMA2-13B. CU.POKer uses
53.7--85.0\% in the configurations that execute successfully.
For LLaMA2-13B at a sequence length of 4k, CU.POKer runs out of memory,
whereas \name achieves 76.8k tokens/s using 98.0\% of the available PEs.

For stencil workloads, \name achieves approximately $1.26\times$ speedup
for Gray-Scott and $1.18$--$1.21\times$ for elastic wave propagation across
the evaluated Z dimensions. Their PE utilization increases from 75.0\% to
96.3\% and from 83.3\% to 97.6\%, respectively. The utilization remains
constant as Z increases, while throughput varies with the amount of work
per PE.

\FloatBarrier
\subsection{Competing with Manually Optimized Systems}
\label{sec:exp:waferllm}

We compare \name with the manually optimized WaferLLM
system~\cite{he2025waferllm}, using its PE task templates as
mapping units. All experiments run on WSE-3 hardware;
configurations labelled WSE-2 restrict the usable region to
the WSE-2 fabric size.

\begin{figure}[!t]
  \centering
  \includegraphics[width=1.00000\linewidth]{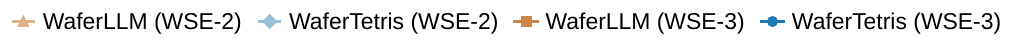}\par\smallskip
  \begin{subfigure}[b]{0.52275\linewidth}
    \centering
    \includegraphics[width=\linewidth]{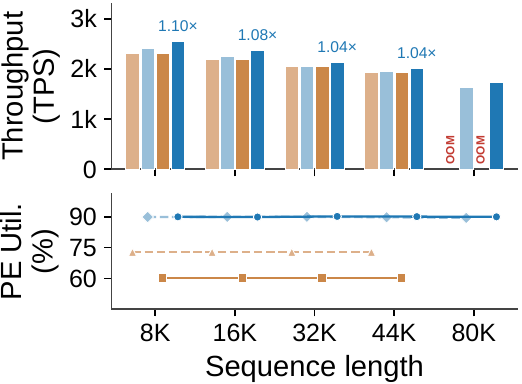}
    \caption{LLaMA3.1-8B}
    \label{fig:waferllm-8b}
  \end{subfigure}\hfill
  \begin{subfigure}[b]{0.45725\linewidth}
    \centering
    \includegraphics[width=\linewidth]{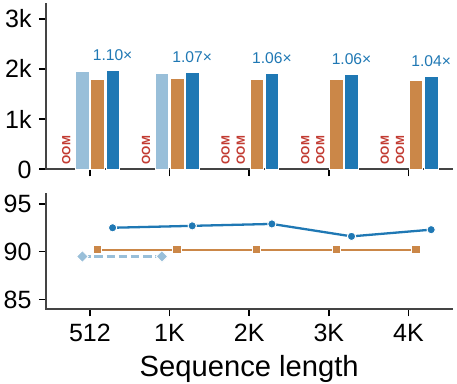}
    \caption{LLaMA2-13B}
    \label{fig:waferllm-13b}
  \end{subfigure}
  \caption{Per-user decoding throughput and PE utilization compared with
  WaferLLM~\cite{he2025waferllm}.}
  \Description{Two panels compare WaferLLM and WaferTetris for LLaMA3.1-8B
  and LLaMA2-13B with WSE-2 and WSE-3 fabric budgets. Each panel contains
  throughput bars and PE-utilization curves. Speedup annotations compare
  the WSE-3 results. Unavailable configurations are marked OOM in red.}
  \label{fig:compared_waferllm}
\end{figure}

\Cref{fig:compared_waferllm} shows that \name achieves
$1.04$--$1.10\times$ WaferLLM's per-user decoding throughput
on both models under the WSE-3 budget wherever both systems run.
Under the WSE-2 budget, \name matches or slightly exceeds
WaferLLM's throughput on LLaMA3.1-8B, demonstrating that
automatic mapping can compete with manually optimized layouts.

\name also supports larger workloads within the same fabric budget.
For LLaMA3.1-8B, the largest successful tested context is 80k
for \name versus 44k for WaferLLM on both budgets, a
$1.8\times$ increase. Under the WSE-2 budget, \name runs
LLaMA2-13B at context lengths of 512 and 1k, while WaferLLM
runs out of memory at every tested length.

\FloatBarrier

\subsection{Effectiveness of FPGA P\&R}
\label{sec:exp:pr-effectiveness}

We evaluate whether FPGA P\&R maintains communication locality in freeform
mappings by comparing their mapping-unit latency with regular rectangle
mappings, which preserve each kernel's native geometry and avoid
freeform-induced path elongation or routing congestion. We use square GEMV and
GEMM kernels on WSE-3 while holding the logical topology, PE allocation, and
kernel implementation fixed: CheckerBoard from the Cerebras CSL
library~\cite{cerebras_gemv_checkerboard_2025}, and WaferLLM's MeshGEMV and
MeshGEMM~\cite{he2025waferllm}.

\begin{figure}[!t]
  \centering
  \includegraphics[width=0.47739\linewidth]{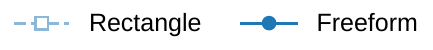}\par\smallskip
  \begin{subfigure}[b]{0.35825\linewidth}
    \centering
    \includegraphics[width=\linewidth]{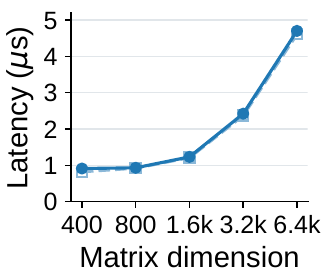}
    \caption{CheckerBoard}
    \label{fig:pr-checkerboard}
  \end{subfigure}\hfill
  \begin{subfigure}[b]{0.31405\linewidth}
    \centering
    \includegraphics[width=\linewidth]{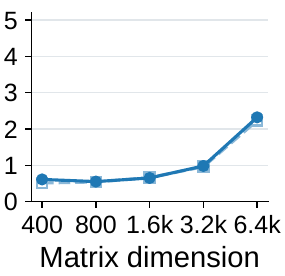}
    \caption{MeshGEMV}
    \label{fig:pr-meshgemv}
  \end{subfigure}\hfill
  \begin{subfigure}[b]{0.30770\linewidth}
    \centering
    \includegraphics[width=\linewidth]{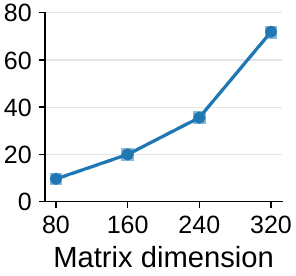}
    \caption{MeshGEMM}
    \label{fig:pr-meshgemm}
  \end{subfigure}
  \caption{Mapping-unit latency under rectangle and freeform P\&R on WSE-3.}
  \Description{Three line-chart panels compare rectangle and freeform mapping
  latency. CheckerBoard and MeshGEMV use square matrix dimensions 400, 800,
  1600, 3200, and 6400; MeshGEMM uses dimensions 80, 160, 240, and 320.
  The curves remain close except at the smallest GEMV configuration.}
  \label{fig:pr-effectiveness}
\end{figure}

For GEMV dimensions from 800 to 6400, freeform latency differs from its
rectangle counterpart by $-1.5$--$+4.1\%$. Across the evaluated MeshGEMM
dimensions from 80 to 320, the difference is $-0.4$--$+1.5\%$.
These results support our insight and justify the efficient rectangle-template
profiling strategy in \Cref{sec:actor_comm_model}: over these ranges,
regular-shape profiles closely approximate freeform mapping latency without
running P\&R for every candidate.

\FloatBarrier

\subsection{Analysis and Ablation Studies}
\label{sec:exp:analysis}
\label{sec:exp:model_accuracy}
We evaluate the accuracy of local latency and system throughput predictions,
and examine whether the model selects island sizes that improve measured
throughput.

\begin{figure}[!t]
  \centering
  \includegraphics[width=0.47739\linewidth]{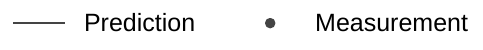}\par\smallskip
  \begin{subfigure}[b]{0.35825\linewidth}
    \centering
    \includegraphics[width=\linewidth]{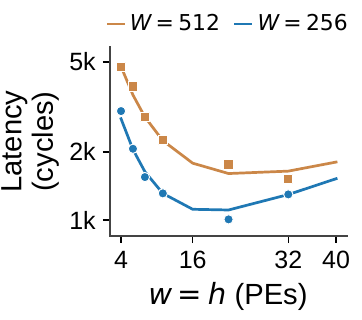}
    \caption{GEMV}
    \label{fig:model:gemv}
  \end{subfigure}\hfill
  \begin{subfigure}[b]{0.31405\linewidth}
    \centering
    \includegraphics[width=\linewidth]{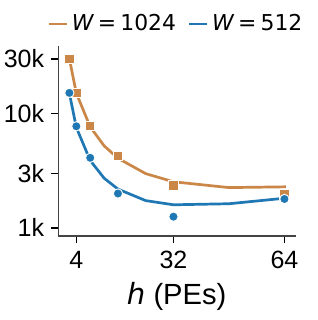}
    \caption{Softmax}
    \label{fig:model:softmax}
  \end{subfigure}\hfill
  \begin{subfigure}[b]{0.30770\linewidth}
    \centering
    \includegraphics[width=\linewidth]{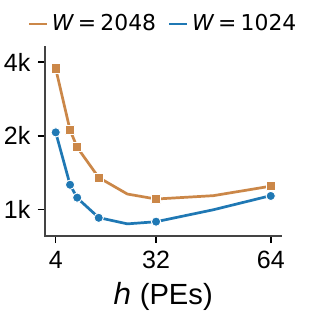}
    \caption{RMSNorm}
    \label{fig:model:rmsnorm}
  \end{subfigure}
  \caption{Predicted and measured local latency across logical island
  sizes for two workloads per kernel. The proposed latency models effectively capture
  the communication bound for larger islands.}
  \Description{Lines show predicted latency and markers show measurements
  for GEMV, Softmax, and RMSNorm. Latency initially decreases as islands
  grow, then flattens or increases. Mean absolute percentage errors over
  the displayed measurements are 4.2, 6.5, and 0.1 percent, respectively.}
  \label{fig:model:latency}
\end{figure}

\begin{figure}[t]
  \centering
  \includegraphics[width=0.74427\linewidth]{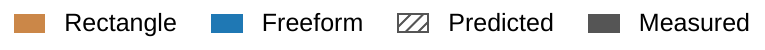}\par\smallskip
  \begin{subfigure}[b]{0.52179\linewidth}
    \centering
    \includegraphics[width=\linewidth]{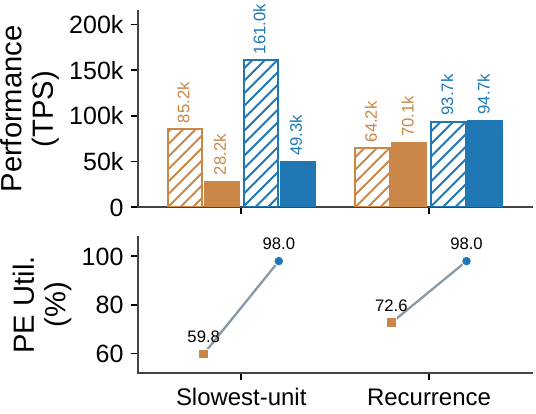}
    \caption{LLaMA3.1-8B (4k)}
    \label{fig:controlled:llama8b}
  \end{subfigure}\hfill
  \begin{subfigure}[b]{0.45821\linewidth}
    \centering
    \includegraphics[width=\linewidth]{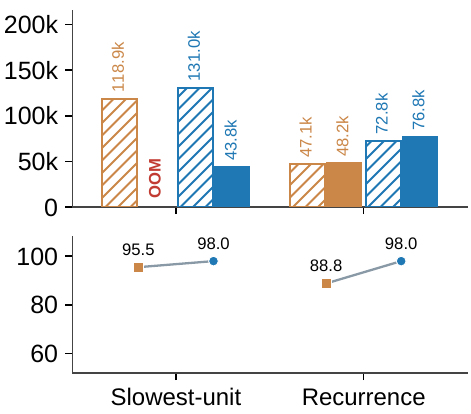}
    \caption{LLaMA2-13B (4k)}
    \label{fig:controlled:llama13b}
  \end{subfigure}
  \caption{Ablation of the performance objective and mapping geometry for two LLMs at a 4k context on WSE-3. The recurrence model improves throughput prediction and mapping selection; freeform mapping further improves measured throughput and PE utilization.}
  \Description{Each model panel shows the two performance objectives, and
  within each objective the rectangular and freeform geometries. Under the
  slowest-unit objective, throughput is predicted at 85.2k and 161.0k tokens
  per second for LLaMA3.1-8B but measured at only 28.2k and 49.3k, while the
  rectangular LLaMA2-13B mapping runs out of memory and the freeform one is
  predicted at 131.0k against a measured 43.8k. Under the recurrence
  objective, the predictions of 64.2k and 93.7k for LLaMA3.1-8B and 47.1k and
  72.8k for LLaMA2-13B match measurements of 70.1k, 94.7k, 48.2k and 76.8k
  within 8.4 percent. Freeform raises PE utilization from 59.8 to 98.0 percent
  and from 72.6 to 98.0 percent on LLaMA3.1-8B, and from 95.5 to 98.0 percent
  and from 88.8 to 98.0 percent on LLaMA2-13B.}
  \label{fig:controlled-ablation}
\end{figure}

\noindent\textbf{Accurate Latency Model.}
\Cref{fig:model:latency} compares the latency model in
\Cref{eq:latency_template} with measurements for GEMV, Softmax, and
RMSNorm, varying logical island size at two workloads per kernel.
For the plotted samples, the model achieves 93.5--99.9\% average prediction
accuracy across the three kernels.
Appendix~\ref{sec:latency_model_cross_validation} reports full-fit and
leave-one-out cross-validation accuracy over all 172 profiled configurations.
It captures both compute-bound and
communication-bound regimes, reproducing the characteristic U-shaped
latency curves. The left side of each curve corresponds to compute-bound
designs, where larger islands reduce work per PE; the right side reflects
communication-bound designs, where communication increasingly limits
performance.

\noindent\textbf{Recurrence Model Finds Better Mappings.}
\Cref{fig:controlled-ablation} compares the two objectives on LLaMA3.1-8B
and LLaMA2-13B at a 4k context. The slowest-unit estimate scores each
mapping unit in isolation and overpredicts throughput by 199--227\%. The
recurrence model instead captures execution and streaming costs along
dependency paths that span multiple units, and
predicts the end-to-end graph within 8.4\%. Modeling the whole graph is what
lets the search distinguish good spatial parallelism assignments from bad
ones: those selected under the recurrence model reach $1.75$--$2.49\times$
the measured throughput of those selected under the slowest-unit estimate.

\noindent\textbf{Freeform Mapping Unlocks Spatial Parallelism.}
\Cref{fig:controlled-ablation} isolates the effect of mapping
geometry by comparing rectangular and freeform layouts under
each performance objective. With the recurrence objective
fixed, freeform mapping increases measured throughput from
70.1k to 94.7k tokens/s on LLaMA3.1-8B and from 48.2k to
76.8k tokens/s on LLaMA2-13B, while raising PE utilization
to 98.0\% on both models. Under the slowest-unit objective,
freeform mapping also improves throughput on LLaMA3.1-8B
and enables LLaMA2-13B to fit in memory.
These results show that relaxing rectangular layout
constraints improves both throughput and workload capacity.

\FloatBarrier

\begin{table}[!t]
  \centering
  \caption{LLaMA2-13B mapping time in minutes. Profiling is shared
  across configurations; per-mapping totals exclude profiling and simulation.}
  \label{tab:exp:mapping_time}
  \begingroup
  \setlength{\tabcolsep}{6pt}
  \renewcommand{\arraystretch}{1.12}
  \footnotesize
  \begin{tabular}{@{}lrrr@{}}
    \toprule
    Stage & 1k & 2k & 4k \\
    \midrule
    Profiling (one-time) & \multicolumn{3}{c}{90.00 (shared)} \\
    \midrule
    Parallelism selection & 0.20 & 0.25 & 0.43 \\
    Placement             & 12.26 & 12.42 & 12.35 \\
    Routing               & 0.50 & 0.58 & 0.30 \\
    CSL compilation       & 23.89 & 23.82 & 24.26 \\
    \midrule
    Per-mapping total     & \textbf{36.84} & \textbf{37.07} & \textbf{37.34} \\
    \bottomrule
  \end{tabular}
  \endgroup
\end{table}

\subsection{Mapping Time Efficiency}
\label{sec:exp:mapping_time}

\name provides an efficient offline mapping flow, completing mapping and
CSL compilation in 36.8--37.3 minutes for the evaluated LLaMA2-13B
configurations (\Cref{tab:exp:mapping_time}). Profiling takes 90 minutes
but is required only once for fixed kernel implementations and a target
platform; subsequent mappings reuse the fitted performance models.
The compiled mappings can also be reused across application runs.

\FloatBarrier

\section{Related Work}

{\bf Automatic Mapping Systems for WSC.}
CU.POKer~\cite{jiang2020cu} and GigaPlacer~\cite{li2021placement},
developed from the ISPD placement contests~\cite{ispd2021contest},
perform automated template-based intra- and inter-island placement
but do not address inter-island routing.
RAM-Wafer~\cite{dai2025ram} uses reinforcement learning for placement
and routing but remains restricted to rectangular templates.
TEMP~\cite{wang2026temp} addresses traffic-aware tensor
partition-mapping at the die level, without customized fine-grained
PE placement or routing.
BusyBarn~\cite{cui2026busybarn} optimizes hierarchical mapping
and fault-tolerant communication for multi-die LLM inference.
\name targets finer-grained PE-level mapping on Cerebras WSE,
explicitly accounting for per-PE memory and router-color constraints.
Sai et al.~\cite{sai2024stencil} automate high-order stencil code
generation and router configuration on the WSE but target a single
application domain rather than general-purpose spatial mapping.
The MLIR pipeline~\cite{stawinoga2026mlir} optimizes within a single
PE island and is orthogonal to our inter-island optimization.
\name jointly explores spatial parallelism, placement, and routing
within and across PE islands for diverse applications while
preserving the provided PE-task code.



{\bf Spatial Dataflow Compilers.}
Spatial dataflow compilers on FPGAs share a similar problem domain with \name, as both map computations onto two-dimensional spatial fabrics. 
At the algorithmic level, frameworks such as ScaleHLS~\cite{ye2022scalehls}, HIDA~\cite{ye2024hida}, AutoDSE~\cite{sohrabizadeh2022autodse}, Allo~\cite{chen2024allo}, and StreamTensor~\cite{ye2025streamtensor} explore design spaces including parallelism, tiling, kernel fusion, and resource allocation for dataflow accelerators. 
However, they cannot address spatial placement and routing, while \name provides an automated approach for the systematic exploration of spatial mapping.
AutoBridge~\cite{guo2021autobridge} combines FPGA placement with dataflow pipelining to achieve high frequency on multi-die FPGAs.
This was later extended by RapidStream~\cite{guo2022rapidstream} for optimizing routing on FPGAs. 
TAPA~\cite{guo2023tapa} further provides a task-parallel programming model with algorithm co-optimization and physical design. 
However, they target coarse-grained mapping across FPGA dies, often up to eight dies per device, while \name presents a fine-grained mapping approach for the mapping exploration at wafer scale.

\section{Discussion and Future Work}
\label{sec:discussion}

\noindent\textbf{Partition scheduling.}
\name starts from a graph of mapping units, with the application choosing
which operators to fuse and which to place on separate islands.
Automating this choice is a natural next step.
Fusion can reduce intermediate data movement, while separating operators
allows independent spatial allocation and pipeline overlap.
A future partition scheduler could generate candidate fusion boundaries
and use \name's performance model to evaluate their memory requirements,
inter-island communication, and pipeline throughput. This would extend
optimization from mapping a given graph to jointly selecting the graph
and its spatial realization.

\noindent\textbf{Further FPGA Optimization.}
\name's architecture model enables the adaptation of FPGA optimizations
beyond placement and routing. This would extend mapping exploration to
transformations of the PE graph, allowing the system to improve communication
locality and resource utilization before physical mapping.
For example, PE replication, inspired by
FPGA register replication~\cite{amd_allow_register_replication_2025}, could
further improve performance by shortening communication paths, as explored
in Appendix~\ref{sec:pe_replication}. A future optimization stage could use
performance estimates to select beneficial transformations under the wafer's
compute, memory, and routing constraints.

\noindent\textbf{Fault and defect exploration.}
Cerebras WSEs use spare cores and reconfigurable, redundancy-enabling couplings
to tolerate manufacturing defects, yet the shipping WSE-3 activates 900,000
of its 970,000 physical cores, leaving roughly 70,000 (7.2\%) inactive%
~\cite{lie2022processor_redundancy,wang2025defect_tolerance}.
As future work, \name could instead consume a fine-grained map of usable PEs
and their post-repair connectivity, encode defects as unavailable placement
sites and routing resources, and place the application around the resulting
holes. Freeform placement could then retain irregular but functional regions
that a rectangular logical array excludes, increasing the usable PE count and
application-level PE utilization.

\section{Conclusions}

This paper presented \name, a system for efficient freeform mapping on
wafer-scale accelerators.
By removing predefined island-shape constraints while preserving PE-task
code and logical connectivity, \name reduces spatial fragmentation and
makes more compute and memory resources available to applications.
It combines a recurrence-based throughput model for spatial-parallelism
selection with a WSE architecture model that enables existing FPGA P\&R
algorithms to generate placements and legal routes.
Our evaluation on Cerebras WSE-3 covers LLM decoding and scientific
stencil workloads.
For LLM decoding, \name achieves $3.24$--$4.42\times$ the system throughput
of the automatic mapping baseline CU.POKer under the same fabric size,
with over 93\% PE utilization.
Compared with WaferLLM, it achieves up to $1.10\times$ the per-user throughput
and supports $1.8\times$ the context length for LLaMA3.1-8B (80k versus 44k).
These results demonstrate that freeform mapping can improve resource
utilization and application performance while reusing existing PE-task
implementations.



\bibliographystyle{ACM-Reference-Format}
\bibliography{sample-base}

\clearpage
\appendix

\section{Cross-Validation of Mapping-Unit Latency Models}
\label{sec:latency_model_cross_validation}

\Cref{fig:model:latency} visualizes a representative subset of the
mapping-unit latency measurements. The complete profiling corpus contains 172
configurations: 88 for GEMV, 36 for Softmax, 30 for RMSNorm, and 18 for
Stencil. The corresponding models fit six, three, three, and four
coefficients, respectively.

We report $R^2$ and mean absolute percentage error (MAPE) on the full-data fit.
To evaluate held-out generalization without selecting a single split, we use
pointwise leave-one-out cross-validation (LOOCV)~\cite{stone1974cross_validation}.
For each kernel with $n$ profiles, LOOCV holds out each profile once, refits
the model on the other $n-1$ profiles, and predicts the held-out latency.
\Cref{tab:latency_model_cross_validation} reports metrics over the resulting
$n$ held-out predictions, including the 95th-percentile absolute percentage
error (P95 APE); no fitting fold failed.

\begin{table}[h]
  \centering
  \caption{Full-fit and held-out accuracy of the mapping-unit latency models.
  $n$ is the number of profiled configurations and $k$ the number of fitted
  coefficients. P95 APE is computed over LOOCV predictions.}
  \label{tab:latency_model_cross_validation}
  \scriptsize
  \setlength{\tabcolsep}{2.2pt}
  \begin{tabular}{@{}lrrccccc@{}}
    \toprule
    & & & \multicolumn{2}{c}{Full fit} & \multicolumn{3}{c}{LOOCV} \\
    \cmidrule(lr){4-5}\cmidrule(l){6-8}
    Operation & $n$ & $k$ & $R^2$ & MAPE & $R^2$ & MAPE & P95 APE \\
    \midrule
    GEMV    & 88 & 6 & 0.999 & 2.1\%  & 0.998 & 2.3\%  & 6.8\%  \\
    Softmax & 36 & 3 & 1.000 & 10.5\% & 1.000 & 11.3\% & 23.0\% \\
    RMSNorm & 30 & 3 & 1.000 & 0.2\%  & 1.000 & 0.2\%  & 0.8\%  \\
    Stencil & 18 & 4 & 0.948 & 10.1\% & 0.898 & 12.8\% & 21.4\% \\
    \bottomrule
  \end{tabular}
\end{table}

LOOCV MAPE ranges from 0.2\% to 12.8\% across the four kernels. The larger P95
errors for Softmax and Stencil show that high $R^2$ does not imply uniformly
small relative error at every configuration. These results evaluate
generalization to held-out points within the profiled corpus; they do not
establish accuracy for unseen kernel families or extrapolation beyond the
profiled ranges.

\FloatBarrier

\section{Beyond P\&R: Optimization via PE Replication}
\label{sec:pe_replication}

\begin{figure}[!t]
  \centering
  \begin{subfigure}[b]{0.48\linewidth}
    \centering
    \includegraphics[width=\linewidth]{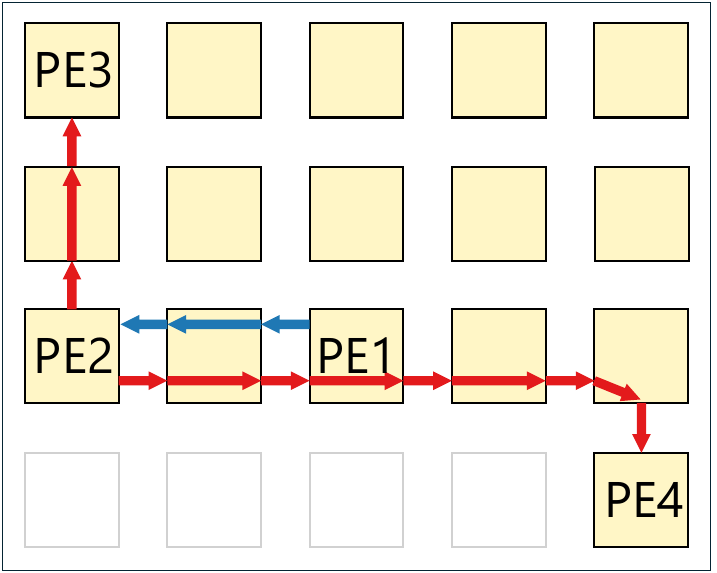}
    \caption{The critical path connects PE2 and PE4.}
    \label{fig:rr:base}
  \end{subfigure}\hfill
  \begin{subfigure}[b]{0.48\linewidth}
    \centering
    \includegraphics[width=\linewidth]{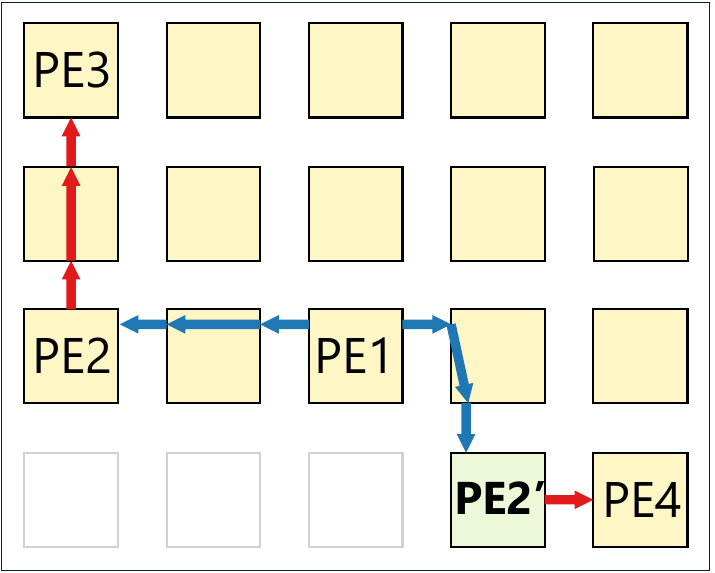}
    \caption{A copy of PE2 is placed near PE4.}
    \label{fig:rr:ourwork}
  \end{subfigure}

  \medskip
  \begin{subfigure}[b]{\linewidth}
    \centering
    \includegraphics[width=\linewidth]{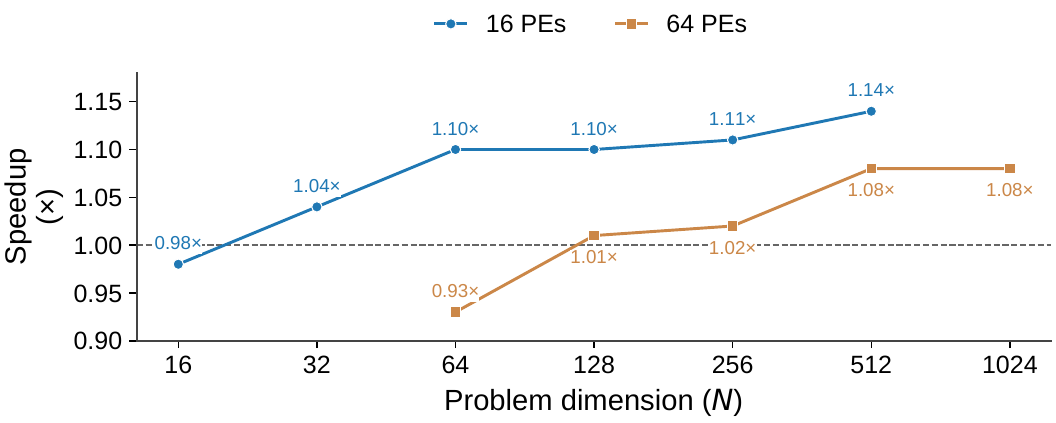}
    \caption{Speedup for an $N\times N$ problem with one additional PE.}
    \label{fig:rr:results}
  \end{subfigure}
  \caption{PE replication as an optimization beyond P\&R. Legend entries
  indicate the original PE counts; replicated mappings use 17 and 65 PEs.
  The dashed line marks unchanged performance.}
  \Description{The first two panels illustrate shortening communication
  by copying PE2 near PE4. The third panel shows the original measurements
  as two lines. Speedup ranges from 0.98 to 1.14 for the 16-PE baseline
  and from 0.93 to 1.08 for the 64-PE baseline. Unreported configurations
  are omitted.}
  \label{fig:rr}
\end{figure}

The architecture model provides an interface for adapting FPGA
optimizations beyond placement and routing to WSE mappings. Register
replication is used in FPGA toolchains to reduce fanout on critical
nets~\cite{scoville2011register,amd_allow_register_replication_2025}.
We adapt this idea as PE replication: duplicate a producer's computation
so that a nearby copy can serve a distant consumer.

\Cref{fig:rr:base} illustrates a mapping in which PE2 broadcasts to PE3
and PE4, with the route to PE4 forming the critical path. In
\Cref{fig:rr:ourwork}, a copy, PE2', is placed near PE4 and performs the
same computation. PE4 can then obtain its input through the shorter
route from PE2'. This transformation trades an additional PE and
duplicated computation for a shorter communication path.

Our current mapping flow does not perform this transformation
automatically. We manually apply it to mappings with 16 and 64 PEs,
adding one PE in each case. \Cref{fig:rr:results} reports speedup over the
corresponding mapping without replication. The 16-PE case reaches
$1.14\times$ at a problem size of $512\times512$; the 64-PE case reaches
$1.08\times$ at $512\times512$ and $1024\times1024$.

Replication is not uniformly beneficial: speedup falls to $0.98\times$
for the smallest reported 16-PE workload and $0.93\times$ for the
smallest reported 64-PE workload. An automated pass must therefore
evaluate the communication benefit against the cost of the extra PE,
duplicated computation, and input delivery. This case study demonstrates
the potential of integrating graph transformations into \name's mapping
flow; selecting and applying such transformations automatically remains
future work.

\FloatBarrier

\end{document}